\documentclass[12pt]{article}
\usepackage{graphicx}
\usepackage{amssymb}
\usepackage{array} 
\usepackage[round]{natbib}
\usepackage[affil-it]{authblk}
\usepackage{bibentry}
\usepackage{amsmath}
\usepackage{booktabs}
\usepackage{dashbox}
\usepackage{float}
\usepackage{placeins}
\usepackage{needspace}
\usepackage{etoolbox}
\usepackage{verbatim}
\usepackage{comment}
\usepackage{color}
\usepackage{makeidx}
\usepackage{bm}
\usepackage{subcaption}
\usepackage{hyperref}
\usepackage{makecell}
\usepackage{setspace}
\usepackage[letterpaper,top=1in,bottom=1in,left=1in,right=1in]{geometry}

\newcommand{\AN}{\mathrm{AN}}
\newcommand{\N}{\mathrm{N}}
\newcommand{\IG}{\mathrm{IG}}

\usepackage{pifont}
\newcommand{\passmark}{{\scriptsize\ding{51}}}
\newcommand{\mcsmark}{\passmark}

\newenvironment{keyword}{%
  \par\noindent\textbf{Keywords }\ignorespaces
}{\par\vspace{1em}}
\newcommand{\kwd}[1]{#1}

\title{Realized Tail-Risk Conditional Autoregressive Expectile \\ Models:  Accurate and Stable Forecasting Performance}

\author[1]{Keisuke Sasaki}
\author[2]{Yasuhiro Omori\thanks{Corresponding author.\\ E-mail addresses: keisukes856@gmail.com (Keisuke Sasaki), omori@e.u-tokyo.ac.jp (Yasuhiro Omori).}}

\affil[1]{Graduate School of Economics, The University of Tokyo}
\affil[2]{Faculty of Economics, The University of Tokyo}

\date{}
\def\spacingset#1{\renewcommand{\baselinestretch}{#1}\small\normalsize}
\DeclareCaptionFont{tablecaption}{%
  \fontsize{10}{11}\selectfont
}
\begin{document}
\spacingset{1}
\maketitle
\begin{abstract}
    A realized tail-risk conditional autoregressive expectile (RTCARE) model is proposed for joint
    forecasting of value-at-risk (VaR) and expected shortfall (ES). Realized tail measures such as the realized quantile and the realized ES, constructed from intraday high-frequency returns, are incorporated into the
    dynamics and measurement equations of the conditional expectile, allowing information about lower-tail risk to enter the forecasts. The parameters are estimated by Markov chain Monte Carlo, and we evaluate the resulting VaR--ES forecasts in numerical examples and in an empirical application using U.S. stock-return data. Across the considered tail levels, the RTCARE models deliver competitive and often superior forecasting performance relative to existing alternatives, with stable results across Monte Carlo replications and rolling forecast evaluations.
\end{abstract}
    \begin{keyword}
    \kwd{Expected shortfall, Expectiles, Markov chain Monte Carlo, Realized quantile, Tail risk, Value-at-risk 
   }
    \end{keyword}
\spacingset{1}
        \section{Introduction}
    
    Value-at-risk (VaR) and expected shortfall (ES) are central tools for measuring financial tail risk. VaR, at probability level $\alpha$,  is the lower $\alpha$-quantile of the
    conditional return distribution and therefore specifies a return threshold
    that is breached with conditional probability $\alpha$. Lower-tail levels
    such as $\alpha=0.01, 0.025$ and $0.05$ are commonly considered in empirical risk
    management.

    ES instead measures the conditional mean return in the tail below VaR and thus quantifies the average severity of losses once the VaR
    threshold has been breached. Unlike VaR, ES accounts for the entire tail and
    is a coherent risk measure as shown in \citet{ArtznerEtAl1999}. Its
    regulatory importance has also increased through the Basel Committee's
    revised market-risk framework, which uses ES in place of VaR for determining
    market-risk capital in Basel I\hspace{-1.2pt}I\hspace{-1.2pt}I \citep[e.g.,][]{BaselCommittee2019}. Joint VaR--ES forecasts are consequently
    of practical and methodological importance. Accurate prediction remains
    challenging, however, since information about extreme
    tails is limited and changes over time with both conditional volatility and
    the shape of the return distribution.

    A common parametric approach is to model conditional volatility using, for
    example, a GARCH model \citep{Engle1982,Bollerslev1986} or a stochastic
    volatility model \citep{Taylor1986}, and then derive VaR and ES under an
    assumed innovation distribution, such as the Gaussian or Student's
    $t$ distribution. The resulting tail forecasts can be sensitive to this
    distributional choice, especially at low probability levels \citep[see, e.g.,][]{TakahashiYamauchiWatanabeOmori26}.
    Semi-parametric alternatives instead model a
    conditional tail functional directly. For example, the conditional autoregressive VaR (CAViaR) model was proposed by \citet{EngleManganelli04} to model the dynamics of conditional quantiles. CAViaR models have been extended to a family of nonlinear dynamic models for estimating time-varying VaR \citep[e.g.,][]{GerlachChenChan11}. 
    Alternatively, expectiles can be used to model the dynamics of VaR and ES. Expectiles are
    defined as minimizers of the expected asymmetric squared loss
    \citep{NewyPowell87}. For a zero conditional mean, the expectile associated
    with a given quantile level has an analytical mapping to both VaR and ES
    \citep{Taylor08}.
    Building on this relationship, \citet{Taylor08} proposed the
    conditional autoregressive expectile (CARE) model, in which the expectile
    evolves dynamically and delivers joint estimation and forecasting of VaR
    and ES without specifying a parametric distribution for returns.
    Markov chain Monte Carlo (MCMC) methods have also been used in related dynamic VaR--ES models to represent parameter uncertainty in tail-risk forecasts \citep[e.g.,][]{ChenGerlachLu12,GerlachChen14}.
    
    Daily returns alone need not exhaust the information available for forecasting
    a daily tail event. High-frequency prices provide realized measures that
    summarize intraday variation. Realized variance is commonly constructed
    as the sum of squared intraday returns and used to estimate daily
    integrated variance
    \citep{AndersenEtAl2001,AndersenEtAl2003}.
    Related realized-volatility models combine daily returns with
    high-frequency volatility measures. Realized GARCH models use a
    measurement equation linking realized volatility to the latent volatility
    state \citep{HansenRGARCH12}, while realized stochastic-volatility models
    incorporate realized volatility into the estimation of a latent
    stochastic-volatility process \citep{Takahashi09,Takahashi23}.
    \citet{GerlachWang22} extended the CARE framework by
    incorporating intraday realized information through a measurement equation
    and lagged realized variables in the expectile dynamics.
    However, realized-volatility measures
    summarize the overall scale of intraday variation and are not designed
    specifically for the lower tail. More recently, \citet{Dimitriadis22}
    proposed realized quantiles (RQ) and realized expected shortfall (RES) as tail-specific,
    daily-scale measures constructed from high-frequency intraday returns. RQ estimates a
    daily return quantile by scaling its intraday counterpart, while RES is
    constructed analogously for ES. They therefore provide direct estimates of
    daily VaR and ES and contain information about the relevant part of the return
    distribution rather than only its overall scale. This motivates using RQ and RES as
    realized tail measures in a dynamic model for joint VaR and ES forecasting.

    This paper proposes realized tail-risk conditional autoregressive expectile
    (RTCARE) models, which extend the CARE framework by incorporating these realized
    tail measures, RQ and RES, into the expectile dynamics and measurement equations
    for joint VaR and ES forecasting.
    Our use of realized measures as observed inputs follows the approach of
    \citet{Takahashi09} and \citet{HansenRGARCH12}, who use realized volatility
    in volatility estimation.
    Furthermore, the proposed RTCARE models can be viewed as
    a tail-risk extension of the realized CARE model of \citet{GerlachWang22},
    which incorporates realized volatility into CARE.
    The proposed models are estimated by MCMC, which
    also allows parameter uncertainty to be incorporated into inference and
    forecasting. 
    We evaluate the forecasting performance of our proposed models using several standard diagnostics. First, we
    use the strictly consistent Fissler--Ziegel loss class \citep{FisslerZiegel16},
    specifically the FZ0 loss proposed by \citet{PattonZiegelChen19}, to
    assess VaR and ES jointly. We then report the coverage and ES diagnostics and
    apply the regression-based ES backtests proposed by
    \citet{BayerDimitriadis22}. Finally, we use the model confidence set (MCS) procedure \citep{HansenLundeNason11} to compare
    predictive accuracy across models and identify models that are not significantly worse than the best model.
    The proposed models are compared with volatility-based, score-driven,
    CARE-type, and realized tail-risk benchmarks in the numerical examples and
    in both the in-sample and one-day-ahead out-of-sample analyses of IBM daily returns.

    The remainder of the paper is organized as follows. Section~2 introduces the
    expectile--VaR--ES relationship and the realized tail-risk measures, and then proposes the RTCARE models. Section~3 describes the MCMC estimation scheme.
    Section~4 presents the numerical examples. Section~5 presents the empirical application using IBM
    data and reports posterior estimates together with in-sample and out-of-sample performance. Section~6 concludes.

    \section{Realized Tail-Risk Conditional Autoregressive Expectile Models}
    
    \subsection{Relationships between Expectiles, VaR, and ES}
    
    {\it Definition of an expectile}.
    \citet{NewyPowell87} introduced expectiles as generalized quantiles defined
    by minimizing an asymmetric quadratic loss and proposed the asymmetric
    least squares (ALS) method for estimation. For a random variable $Y$ and $\tau\in(0,1)$, the $\tau$-level
    expectile is defined by the following optimization problem.
    \begin{align}
        \min_{\mu} \mathbb{E}\left[ |\tau - I(Y < \mu)| (Y - \mu)^2 \right],
    \end{align}
    where $I(\cdot)$ is the indicator function. Let $\mu_t(\tau)$ denote the conditional $\tau$-level expectile of $y_t$ given the information set $\mathcal{F}_{t-1}$. The ALS criterion for $n$ observations is
    \begin{equation}
        \sum_{t=1}^{n} |\tau - I(y_t < \mu_t(\tau))| (y_t - \mu_t(\tau))^2.
        \label{eq:ALSsample}
    \end{equation}

    \noindent
    {\it Definitions of VaR and ES}.  The value-at-risk at level $\alpha$ ($\alpha\in(0,1)$) is defined as $\mathrm{VaR}(\alpha) := \inf\{y\in\mathbb{R} \mid F(y) \geq \alpha\}$,
    where $F$ denotes the cumulative distribution function of $Y$. The
    expected shortfall at level $\alpha$ ($\alpha\in(0,1)$) is defined as $\mathrm{ES}(\alpha) := \mathbb{E}\big[Y \mid Y \leq \mathrm{VaR}(\alpha)\big]$.
    
    \noindent
    {\it Relationships between expectiles, VaR, and ES}.
    For the expectile level $\tau$ corresponding to quantile level $\alpha$, $\mu_t(\tau)=q_t(\alpha)$ \citep{Jones1994}, and \citet{Taylor08} gives
    \begin{align} 
        \label{eq:expectile-es-map}
        ES_{t, \alpha} = \left(1 + \frac{\tau}{(1 - 2\tau)\alpha}\right)\mu_t(\tau)
        -\frac{\tau}{(1 - 2\tau)\alpha}\mathbb{E}(y_t\mid\mathcal{F}_{t-1}).
    \end{align}
    For continuous conditional return distributions, $\alpha=\Pr(y_t<\mu_t(\tau)\mid\mathcal{F}_{t-1})$. We impose a zero conditional mean, eliminating the expected-return term, and select $\tau$ by matching empirical coverage to $\alpha$ using the grid search in Section~\ref{sec:model}.
    \subsection{Realized Tail-Risk Measures for VaR and ES}
    \label{sec:realized-measures}
    
    High-frequency intraday data provide realized measures that add information
    about daily tail risk. \citet{Dimitriadis22} introduced realized quantiles
    (RQ) and realized expected shortfall (RES) as daily tail-risk measures
    constructed from intraday returns. RQ is an intraday-based estimate of a
    lower-tail VaR, while RES is the corresponding estimate of ES. In this
    paper, we refer to RQ and RES jointly as realized tail-risk measures.

    Because intraday volatility varies over the trading day, a fixed
    calendar-time grid does not allocate observations evenly across periods of
    high and low activity. We therefore use business-time sampling (BTS),
    following the business-time construction of \citet{DongTse17}, to account
    for this intraday variation. The detailed construction is given in
    the Supplementary Material, Section~A.
    
    \subsection{Realized Tail-Risk CARE Models}
    \label{sec:model}
     The realized tail-risk conditional autoregressive expectile (RTCARE) models extend the CARE model introduced by \citet{Taylor08}. We propose two RTCARE specifications, RTCARE and RTCARE-RES, which utilize the realized tail-risk measures as described below.\\
    \noindent
    {\bf RTCARE model}. 
    First, for a fixed $\tau\in(0,1)$, the RTCARE model is based on RQ and defined as 
    \begin{align}
        y_t &= \mu_t+\varepsilon_t, \qquad \varepsilon_t\sim\AN(0,\sigma_r^2,\tau),
        \label{eq:rcare-rq-return}\\
        \mu_t &= \beta_0+\beta_1\mu_{t-1}+\beta_2 RQ_{t-1},
        \label{eq:rcare-rq-recursion}\\
        RQ_t &= \alpha_q+\beta_q\mu_t+u_t, \qquad u_t\sim\N(0,\sigma_u^2),
        \label{eq:rcare-rq-measurement}
    \end{align}
    where AN denotes the asymmetric normal distribution.
    Following \citet{GerlachWang22}, we use the asymmetric normal distribution as a pseudo-likelihood device.
    Rather than asserting that returns literally follow this distribution, it is chosen so that
    the likelihood contribution takes the form of the expectile loss.  Under
    this pseudo-likelihood, the contribution of (\ref{eq:rcare-rq-return}) has the kernel
    \begin{equation}
      p(y_t\mid\mu_t,\sigma_r^2,\tau)\propto(\sigma_r^2)^{-1/2}
        \exp\left\{-\frac{|\tau-I(y_t<\mu_t)|(y_t-\mu_t)^2}{\sigma_r^2}\right\}.
        \label{eq:an-kernel}
    \end{equation}
    For fixed $\tau$ and $\sigma_r^2$, maximizing this return pseudo-likelihood is equivalent to minimizing the sample analogue
    of the expectile loss in \eqref{eq:ALSsample}.
    Equation~\eqref{eq:rcare-rq-recursion} specifies the transition dynamics
    of the latent expectile through its own lag and lagged RQ.
    
    Moreover, the measurement equation (\ref{eq:rcare-rq-measurement}) links the latent daily expectile to the observed RQ through measurement error. Although we estimate $\alpha_q$ to take account of the possible bias in the RQ, we could set $\alpha_q=0$. We shall investigate whether such a restriction improves tail-risk forecasting performance in numerical examples and empirical studies.
    Substituting the measurement equation into the expectile recursion yields
    the implied feedback coefficient $\beta_1+\beta_2\beta_q$.  We therefore
    impose the following restriction on this coefficient.
    \begin{equation}
        \left|\beta_1+\beta_2\beta_q\right|<1.
        \label{eq:restriction}
    \end{equation}
    \noindent
    To select $\tau$, we use a grid search. For each candidate
    value, we estimate the proposed model by MCMC and use the
    posterior median of $\mu_t$. We choose the candidate for which the empirical
    violation rate
    $n^{-1}\sum_{t=1}^n I(y_t < \mu_t)$ is closest to the target coverage
    level $\alpha$. \vspace{3mm}\\
    \noindent
    {\bf RTCARE-RES model}. Second, we extend the RTCARE model to incorporate RES, using the relationship in (\ref{eq:expectile-es-map}). For this specification, the expectile recursion in
    \eqref{eq:rcare-rq-recursion} is augmented with lagged RES:
    \begin{equation}
        \mu_t=\beta_0+\beta_1\mu_{t-1}+\beta_2RQ_{t-1}+\beta_3RES_{t-1}.
        \label{eq:rcare-res-recursion}
    \end{equation}
    The return and RQ measurement equations are unchanged, and the RES measurement equation is
    \begin{align}
    RES_t &= \alpha_e+\beta_e\kappa \mu_t+w_t, \qquad w_t\sim\N(0,\sigma_w^2),
    \label{eq:rcare-res-measurement}\\
    & \hspace{5mm}\kappa = 1 + \frac{\tau}{(1-2\tau)\alpha}. \nonumber
    \end{align}
    Substitution of the two measurement equations gives the implied feedback coefficient
    $\beta_1+\beta_2\beta_q+\beta_3\beta_e\kappa$; we impose the corresponding coefficient restriction
    $\left|\beta_1+\beta_2\beta_q+\beta_3\beta_e\kappa\right|<1$. Using RES computed from intraday returns may improve tail-risk forecasting performance. We also compare this specification with RTCARE models that do not include RES in the numerical examples and empirical application. The selection of $\tau$ follows the procedure above.
    
    \section{MCMC Algorithm for the RTCARE Model}
    \label{sec:bayes}
    Define $\theta=(\beta_0,\beta_1,\beta_2, \alpha_q,\beta_q,\sigma_r^2,\sigma_u^2)'$, $y=(y_1,\ldots,y_n)'$, and $x=(x_1,\ldots,x_n)'$, where $x_t=RQ_t$.
    Before imposing the coefficient restriction, we specify independent proper priors as follows.
    $$
    \beta_0,\beta_1,\beta_2,\alpha_q,\beta_q
    \sim \mathcal{N}(0,10),
    \qquad
    \sigma_r^2,\sigma_u^2
    \sim \mathcal{IG}(0.02,0.02).
    $$
    The coefficient restriction in \eqref{eq:restriction} is imposed on the
    coefficient draws. 
    For fixed $\tau$, posterior sampling uses a Metropolis-within-Gibbs
    algorithm. The MCMC algorithm proceeds as follows.\\
    \begin{enumerate}
        \item Initialize
        $(\beta_0,\beta_1,\beta_2,\alpha_q,\beta_q,\sigma_r^2,\sigma_u^2)$
        within the parameter support and compute the corresponding expectile
        path.
        \item Sample
        $(\alpha_q,\beta_0,\beta_1,\beta_2,\beta_q)\mid
        \sigma_r^2,\sigma_u^2,y,x,\tau$ as follows.
        \begin{enumerate}
            \item Integrate out $\alpha_q$ and sample $(\beta_0,\beta_1,\beta_2,\beta_q)\mid
            \sigma_r^2,\sigma_u^2,y,x,\tau$ by a joint random-walk
            Metropolis--Hastings step.
            \item Sample $\alpha_q\mid\beta_0,\beta_1,\beta_2,\beta_q,
            \sigma_u^2,y,x,\tau$ from its normal full conditional
            distribution.
        \end{enumerate}
        \item Sample $\sigma_r^2$ from its inverse-gamma full conditional
        distribution.
        \item Sample $\sigma_u^2$ from its inverse-gamma full conditional
        distribution.
        \item Repeat Steps 2--4 for a sufficiently large number of iterations.
    \end{enumerate}
    The details of the sampling algorithm are given in the Supplementary Material,
    Section~B.
    \section{Illustrative Numerical Examples}
    \label{sec:numerical-examples}

    \subsection{Data-Generating Process for the Numerical Examples}
    
    To assess finite-sample performance, we generate 50 independent
    replications from the data-generating process described below.  Each
    replication uses a 1000-day rolling estimation window and produces 1000
    one-step-ahead forecasts.  Every day contains $c=130$ intraday returns,
    corresponding to three-minute sampling over a 390-minute trading day.
    We use this frequency because the three-minute interval produced the most
    stable RQ and RES series among the sampling intervals considered for the
    U.S. stock returns in the empirical application.
    We evaluate the resulting forecasts using the criteria defined below.
    
    The daily return $y_t$ is assumed to have zero conditional mean and conditional variance
    following the GJR--GARCH recursion \citep{GJR93}
    \begin{equation}
      y_t=\sqrt{h_t}\,\varepsilon_t,\qquad
      h_t=\omega+\{0.04+0.10I(y_{t-1}<0)\}y_{t-1}^2
      +0.88h_{t-1},
      \label{eq:numerical-gjr}
    \end{equation}
    where $\omega$ is selected to set the unconditional variance to the
    prescribed target level.  The standardized innovation is a skewed
    Student-$t$ variate with six degrees of freedom. The standardized
    innovation is defined by
    \begin{equation}
      \varepsilon_t=\frac{X_t-\mathbb E(X_t)}
      {\sqrt{\operatorname{Var}(X_t)}},
      \qquad
      X_t=\{1.35I(T_t<0)+0.85I(T_t\geq0)\}T_t,
      \qquad T_t\sim t_6.
      \label{eq:numerical-innovation}
    \end{equation}
    This specification introduces a leverage effect and a heavier left tail.

    Each simulated day is represented by $c=130$ three-minute returns.
    Provisional intraday returns are first generated from
    $r^{\star}_{t,j}\mid h_t \mathrel{\stackrel{\mathrm{iid}}{\sim}}
    \mathcal{N}\left(0,\frac{h_t}{c}\right)$ for $j=1,\ldots,c$.
    Because the provisional returns will not generally sum to the generated
    daily return $y_t$, the discrepancy is distributed equally across the
    $c$ intraday returns. The recentered returns
    are $r_{t,j}=r^{\star}_{t,j}
    +\frac{y_t-\sum_{\ell=1}^{c}r^{\star}_{t,\ell}}{c}$, with
    $\sum_{j=1}^{c}r_{t,j}=y_t$. The realized-variance measure $RV_t$ is
    then computed from the recentered returns as
    $RV_t=\sum_{j=1}^{c}r_{t,j}^{\,2}$.
    The RQ and RES are computed from the intraday returns
    $\{r_{t,j}\}_{j=1}^{c}$ using the empirical lower-tail quantile and
    tail-average definitions described in the Supplementary Material, Section A.

    \subsection{Three RTCARE Models and Ten Competing Models}
    \label{sec:competing-models}
    
   In the numerical example, we compare three variants of the proposed RTCARE models with ten competing models in terms of VaR and ES forecasting performance. The proposed RTCARE models consist of the baseline RTCARE specification, RTCARE NO INT, which sets the intercept in the RQ measurement equation to zero ($\alpha_q=0$), and RTCARE-RES, which adds lagged RES to the expectile recursion and a second measurement equation for RES in parallel with the RQ equation. The models included in the comparison are listed below.\\

\noindent
 {\bf RTCARE models}:
 \begin{enumerate}
     \item RTCARE
     \item RTCARE NO INT (RTCARE with no intercept in the RQ measurement equation, {\it i.e.}, $\alpha_q \equiv 0$)
     \item RTCARE-RES
 \end{enumerate}
 
  \noindent
   The ten competing models fall into four classes. They are the CARE RV family, the generalized autoregressive score (GAS) model, the GARCH family, and the HAR model. Details are given below. \\ 

   \noindent
   {\bf Competing models}:\\

   \noindent
    (1) {\it CARE RV family models.}  We consider three CARE models combined with various realized measures: (i) RV, (ii) RV and RQ, and (iii) RV, RQ and RES. The first CARE model with RV (denoted as CARE RV) is  considered in \citet{GerlachWang22}, and we extend it to include lagged realized tail-risk measures in the expectile transition in other specifications. That is, one specification uses RV and RQ (denoted as CARE RV-RQ), whereas the other adds RES as an additional realized tail-risk regressor (denoted as CARE RV-RQ-RES), without a separate measurement equation. 
    For the realized CARE class, we consider the Realized CARE-SAV specification proposed in \citet{GerlachWang22}. It combines the expectile dynamics with the realized volatility information and a separate equation for the corresponding realized measure. The detailed specification is given in the Supplementary Material, Section~C.\\

   \noindent
   (2) {\it One-factor GAS model.} We consider the one-factor generalized autoregressive score (GAS) model proposed by \citet{PattonZiegelChen19}, which dynamically updates the parameter governing conditional VaR and ES using the score of the FZ0 loss. FZ0 is the particular formulation proposed by \citet{PattonZiegelChen19} within the strictly consistent Fissler--Ziegel loss class introduced by \citet{FisslerZiegel16}. The parameters are estimated by frequentist M-estimation using the FZ0 loss. The definition of FZ0 loss is given in Section~\ref{sec:evaluation-criteria}.\\

   \noindent
   (3) {\it GARCH family models.} The four volatility-based competitors are GARCH(1,1), GARCH-t(1,1) \citep{Bollerslev1986,TimtGARCH87}, Realized GARCH(1,1), and Realized GARCH-t(1,1) \citep{HansenRGARCH12}. Their VaR and ES forecasts are obtained from the estimated conditional return volatility and the assumed innovation distribution, with parameters estimated by MCMC.\\

    \noindent
    (4) {\it HAR models for RQ and RES}. The heterogeneous autoregressive (HAR) model was originally introduced as a parsimonious model for realized-volatility
    dynamics and is widely used for volatility forecasting \citep{Corsi2009}. Following \citet{Dimitriadis22}, we use HAR-type dynamics to construct one-step-ahead forecasts of RQ and RES. We refer to this specification as the HAR-RQ-RES model. It forecasts the RQ and the RES using daily, weekly, and monthly autoregressive components. \\
   
    \noindent
    Table~\ref{tab:comparison_models} summarizes the ten competing models by the model class and acronym.
    The full equations, distributional assumptions, and VaR--ES mappings for all competing models are provided in the Supplementary Material, Section~C.
    \vspace{10mm}
\begin{table}[htbp]
    \centering
    \captionsetup{
  font=tablecaption,
  skip=3pt
}
    \caption{Ten competing models.}
    \label{tab:comparison_models}
    \fontsize{10}{11}\selectfont
    \setlength{\tabcolsep}{3pt}
    \renewcommand{\arraystretch}{0.95}
    \begin{tabular}{@{}ll@{}}
        \toprule
        Model class & Model acronym \\
        \midrule
        CARE RV family 
                     & CARE RV \\
                     & CARE RV-RQ \\
                     & CARE RV-RQ-RES \\
                     & Realized CARE-SAV \\
        \midrule
        GAS          & One-factor GAS \\
        \midrule
        GARCH family & GARCH \\
                     & GARCH-t \\
                     & Realized GARCH \\
                     & Realized GARCH-t \\
        \midrule
        HAR          & HAR-RQ-RES \\
        \bottomrule
    \end{tabular}
\end{table}
    \subsection{Evaluation Criteria for Backtesting}
    \label{sec:evaluation-criteria}
    We evaluate VaR and ES estimates and forecasts using several quantitative
    criteria. The definitions and implementation of each criterion are given below.\\

    \noindent
    {\it FZ0 loss function}. VaR and ES are jointly elicitable, meaning that their
    true pair uniquely minimizes the expected value of a single scoring function.
    The FZ loss functions form a class of strictly consistent
    joint loss functions for these two quantities \citep{FisslerZiegel16}. Under
    the convention that both VaR and ES are negative, \citet{PattonZiegelChen19}
    define FZ0 as the particular member of this class whose loss differences are
    homogeneous of degree zero. This homogeneity makes loss comparisons
    invariant to a common positive rescaling of returns and the corresponding VaR and ES forecasts. For an
    observed return $y_t$, let $q_t$ and $e_t<0$ denote the one-step-ahead VaR
    and ES forecasts for date $t$. The FZ0 loss is
    \begin{equation}
     L_{FZ0}(y_t,q_t,e_t, \alpha)=
     -\frac{I(y_t\leq q_t)(q_t-y_t)}{\alpha e_t}
     +\frac{q_t}{e_t}+\log(-e_t)-1.
     \label{eq:fz0}
    \end{equation}
    The sample mean of this loss is computed over the relevant evaluation
    period. Smaller values of the FZ0 loss are considered to indicate more
    accurate joint VaR--ES forecasts.\\

    \noindent
    {\it $D$ diagnostic.} The $D$ diagnostic of
    \citet{EmbrechtsKaufmannPatie05} evaluates the accuracy of ES forecasts by
    measuring the discrepancy between realized lower-tail returns and their
    corresponding ES forecasts. For $H$ one-step-ahead forecasts, the
    diagnostic is computed from $\delta_t=y_t-e_t$.
    Let $c_\alpha$ be the empirical $\alpha$-quantile of
    $\{\delta_t\}_{t=1}^H$. Then
    \begin{equation}
     D_1=
     \frac{\sum_{t=1}^{H}I(y_t<q_t)\delta_t}
          {\sum_{t=1}^{H}I(y_t<q_t)},\qquad
     D_2=
     \frac{\sum_{t=1}^{H}I(\delta_t<c_\alpha)\delta_t}
          {\sum_{t=1}^{H}I(\delta_t<c_\alpha)},\qquad
     D=\frac{|D_1|+|D_2|}{2}.
     \label{eq:d-diagnostic}
    \end{equation}
    Smaller $D$ indicates a smaller average discrepancy between realized tail
    returns and the corresponding ES values.\\

    \noindent
    {\it Violation rate}. The violation rate evaluates the calibration of the VaR values alone,
    \begin{equation}
     \widehat v=H^{-1}\sum_{t=1}^H I(y_t\leq q_t),
     \label{eq:violation-rate}
    \end{equation}
    and is compared with the nominal level $\alpha$.\\

    \noindent
    {\it Root mean squared error (RMSE).} In the numerical example, the data-generating process
    provides the true conditional VaR and ES. The RMSE is therefore computed
    relative to these true values.

    \subsection{Comparison of Predictive Performance}
    Tables~\ref{tab:sim-gjr-fz0}, \ref{tab:sim-gjr-d}, and
    \ref{tab:sim-gjr-violation} report the replication means and standard
    deviations for the three main evaluation criteria at all four quantile levels, $\alpha = 0.05, 0.025, 0.01$ and 0.005.
    The VaR and ES RMSE values are reported separately in
    the Supplementary Material, Section~D.
    \vspace{3mm}

\begin{table}[htbp]
\centering
\captionsetup{
  font=tablecaption,
  skip=3pt,
  width=0.85\textwidth
}
\caption{Numerical example. Mean FZ0 losses across the four tail levels. Entries are replication means with standard deviations in parentheses. * denotes the best entry in each column.}
\label{tab:sim-gjr-fz0}
\fontsize{10}{11}\selectfont
\setlength{\tabcolsep}{2pt}
\renewcommand{\arraystretch}{0.88}
\begin{tabular*}{0.85\textwidth}{@{\extracolsep{\fill}}lrrrr}
\toprule
Model & $5\%$ & $2.5\%$ & $1\%$ & $0.5\%$ \\
\midrule
RTCARE & 1.374 (0.184) & 1.574 (0.195) & 1.821 (0.227) & 2.013 (0.276) \\
RTCARE NO INT & *1.363 (0.184) & *1.556 (0.192) & *1.792 (0.213) & *1.962 (0.246) \\
RTCARE-RES & 1.375 (0.184) & 1.574 (0.193) & 1.819 (0.222) & 2.010 (0.266) \\
\midrule
CARE RV & 1.385 (0.185) & 1.588 (0.200) & 1.833 (0.234) & 2.005 (0.276) \\
CARE RV-RQ & 1.405 (0.204) & 1.622 (0.223) & 1.956 (0.346) & 2.337 (0.803) \\
CARE RV-RQ-RES & 1.419 (0.205) & 1.653 (0.250) & 2.014 (0.394) & 2.438 (1.701) \\
Realized CARE-SAV & 1.377 (0.185) & 1.577 (0.197) & 1.822 (0.230) & 1.998 (0.274) \\
\midrule
One-factor GAS & 1.387 (0.193) & 1.614 (0.221) & 1.944 (0.345) & 2.257 (0.393) \\
\midrule
GARCH & 1.387 (0.197) & 1.623 (0.226) & 1.973 (0.297) & 2.317 (0.400) \\
GARCH-t & 1.380 (0.196) & 1.581 (0.209) & 1.815 (0.231) & 1.976 (0.271) \\
Realized GARCH & 1.388 (0.200) & 1.625 (0.227) & 1.979 (0.296) & 2.323 (0.397) \\
Realized GARCH-t & 1.379 (0.197) & 1.581 (0.207) & 1.816 (0.232) & 1.967 (0.269) \\
\midrule
HAR-RQ-RES & 1.393 (0.208) & 1.656 (0.250) & 2.087 (0.353) & 2.413 (0.470) \\
\bottomrule
\end{tabular*}
\end{table}
\vspace{3mm}
\begin{table}[htbp]
\centering
\captionsetup{
  font=tablecaption,
  skip=3pt,
  width=0.85\textwidth
}
\caption{Numerical example. $D$ diagnostics across the four tail levels. Entries are replication means with standard deviations in parentheses. Undefined $D$ values are omitted from the corresponding mean and standard deviation. * denotes the best entry in each column.}
\label{tab:sim-gjr-d}
\fontsize{10}{11}\selectfont
\setlength{\tabcolsep}{2pt}
\renewcommand{\arraystretch}{0.88}
\begin{tabular*}{0.85\textwidth}{@{\extracolsep{\fill}}lrrrr}
\toprule
Model & $5\%$ & $2.5\%$ & $1\%$ & $0.5\%$ \\
\midrule
RTCARE & 0.203 (0.137) & 0.288 (0.215) & 0.535 (0.353) & 0.816 (0.609) \\
RTCARE NO INT & *0.180 (0.133) & 0.289 (0.216) & *0.443 (0.293) & 0.684 (0.477) \\
RTCARE-RES & 0.194 (0.125) & *0.283 (0.202) & 0.509 (0.350) & 0.746 (0.575) \\
\midrule
CARE RV & 0.186 (0.117) & 0.284 (0.216) & 0.495 (0.371) & 0.743 (0.621) \\
CARE RV-RQ & 0.250 (0.231) & 0.383 (0.350) & 0.803 (0.524) & 1.470 (1.293) \\
CARE RV-RQ-RES & 0.275 (0.248) & 0.440 (0.382) & 0.930 (0.816) & 1.234 (1.125) \\
Realized CARE-SAV & 0.199 (0.129) & 0.291 (0.213) & 0.519 (0.371) & 0.757 (0.608) \\
\midrule
One-factor GAS & 0.221 (0.184) & 0.350 (0.285) & 0.946 (1.261) & 1.520 (1.240) \\
\midrule
GARCH & 0.674 (0.321) & 0.905 (0.429) & 1.304 (0.640) & 1.598 (0.917) \\
GARCH-t & 0.378 (0.233) & 0.426 (0.301) & 0.575 (0.479) & 0.642 (0.667) \\
Realized GARCH & 0.677 (0.308) & 0.920 (0.420) & 1.328 (0.640) & 1.628 (0.906) \\
Realized GARCH-t & 0.349 (0.206) & 0.415 (0.286) & 0.524 (0.494) & *0.631 (0.661) \\
\midrule
HAR-RQ-RES & 0.763 (0.356) & 1.090 (0.470) & 1.669 (0.752) & 2.199 (1.142) \\
\bottomrule
\end{tabular*}
\end{table}
\clearpage
\begin{table}[htbp]
\centering
\captionsetup{
  font=tablecaption,
  skip=3pt,
  width=0.85\textwidth
}
\caption{Numerical example. Violation rates (in percent) across the four tail levels. Entries are replication means with standard deviations in parentheses. * denotes the best entry in each column.}
\label{tab:sim-gjr-violation}
\fontsize{10}{11}\selectfont
\setlength{\tabcolsep}{2pt}
\renewcommand{\arraystretch}{0.88}
\begin{tabular*}{0.85\textwidth}{@{\extracolsep{\fill}}lrrrr}
\toprule
Model & $5\%$ & $2.5\%$ & $1\%$ & $0.5\%$ \\
\midrule
Nominal rate & 5.000 & 2.500 & 1.000 & 0.500 \\
\midrule
RTCARE & 5.058 (0.661) & *2.502 (0.441) & 1.052 (0.373) & 0.570 (0.267) \\
RTCARE NO INT & 5.120 (0.533) & 2.530 (0.375) & 1.084 (0.276) & *0.536 (0.235) \\
RTCARE-RES & *4.990 (0.634) & 2.514 (0.445) & 1.064 (0.350) & 0.590 (0.260) \\
\midrule
CARE RV & 5.048 (0.672) & 2.506 (0.443) & *1.032 (0.345) & 0.548 (0.273) \\
CARE RV-RQ & 5.166 (0.738) & 2.646 (0.570) & 1.202 (0.426) & 0.752 (0.422) \\
CARE RV-RQ-RES & 5.208 (0.801) & 2.682 (0.584) & 1.256 (0.461) & 0.832 (0.428) \\
Realized CARE-SAV & 5.038 (0.712) & 2.528 (0.468) & 1.048 (0.374) & 0.590 (0.284) \\
\midrule
One-factor GAS & 5.420 (0.682) & 2.884 (0.551) & 1.266 (0.392) & 0.718 (0.295) \\
\midrule
GARCH & 6.098 (0.792) & 3.982 (0.599) & 2.326 (0.469) & 1.672 (0.377) \\
GARCH-t & 6.608 (0.870) & 3.804 (0.546) & 1.678 (0.333) & 0.976 (0.274) \\
Realized GARCH & 5.950 (0.777) & 3.896 (0.629) & 2.292 (0.452) & 1.672 (0.382) \\
Realized GARCH-t & 6.472 (0.779) & 3.656 (0.576) & 1.630 (0.334) & 0.906 (0.287) \\
\midrule
HAR-RQ-RES & 6.044 (0.805) & 4.140 (0.735) & 2.548 (0.565) & 1.602 (0.397) \\
\bottomrule
\end{tabular*}
\vspace{3mm}
\end{table}

\begin{figure}[htbp]
      \centering
      \makebox[\textwidth][c]{%
        \includegraphics[width=\textwidth,height=0.78\textheight,keepaspectratio]{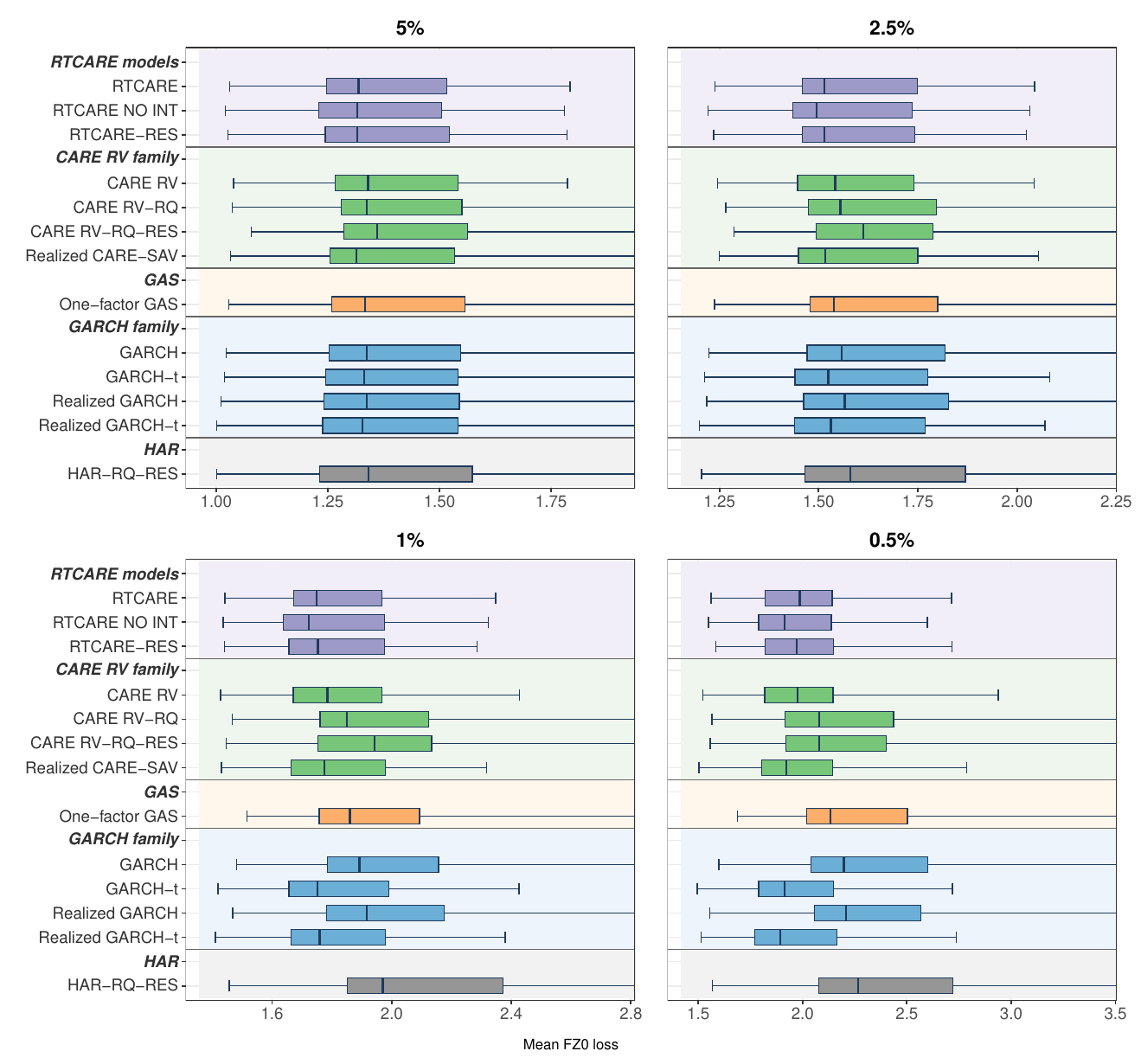}}
      \caption{Boxplots of mean FZ0 loss across the 50 replications. Boxes show the interquartile range, the center line is the median, and whiskers show the observed range.}
      \label{fig:numerical-fz0}
    \vspace{6mm}
    \end{figure}

\begin{figure}[htbp]
      \centering
      \makebox[\textwidth][c]{\includegraphics[width=\textwidth,height=0.78\textheight,keepaspectratio]{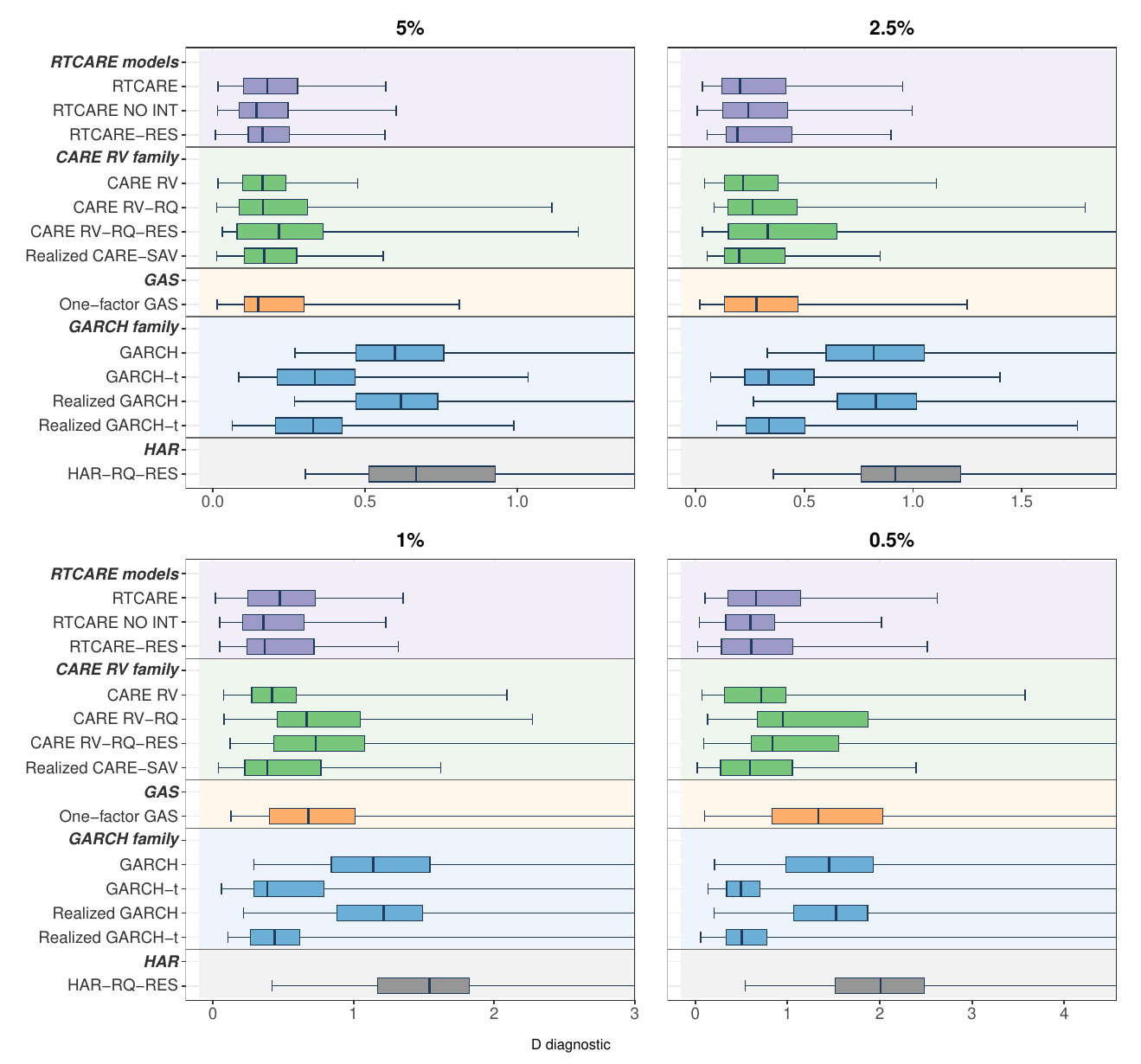}}
      \caption{Boxplots of the $D$ diagnostic across the 50 replications. Boxes show the interquartile range, the center line is the median, and whiskers show the observed range.}
      \label{fig:numerical-d}
      \vspace{6mm}
    \end{figure}

\begin{figure}[H]
      \centering
      \makebox[\textwidth][c]{%
        \includegraphics[width=\textwidth,height=0.78\textheight,keepaspectratio]{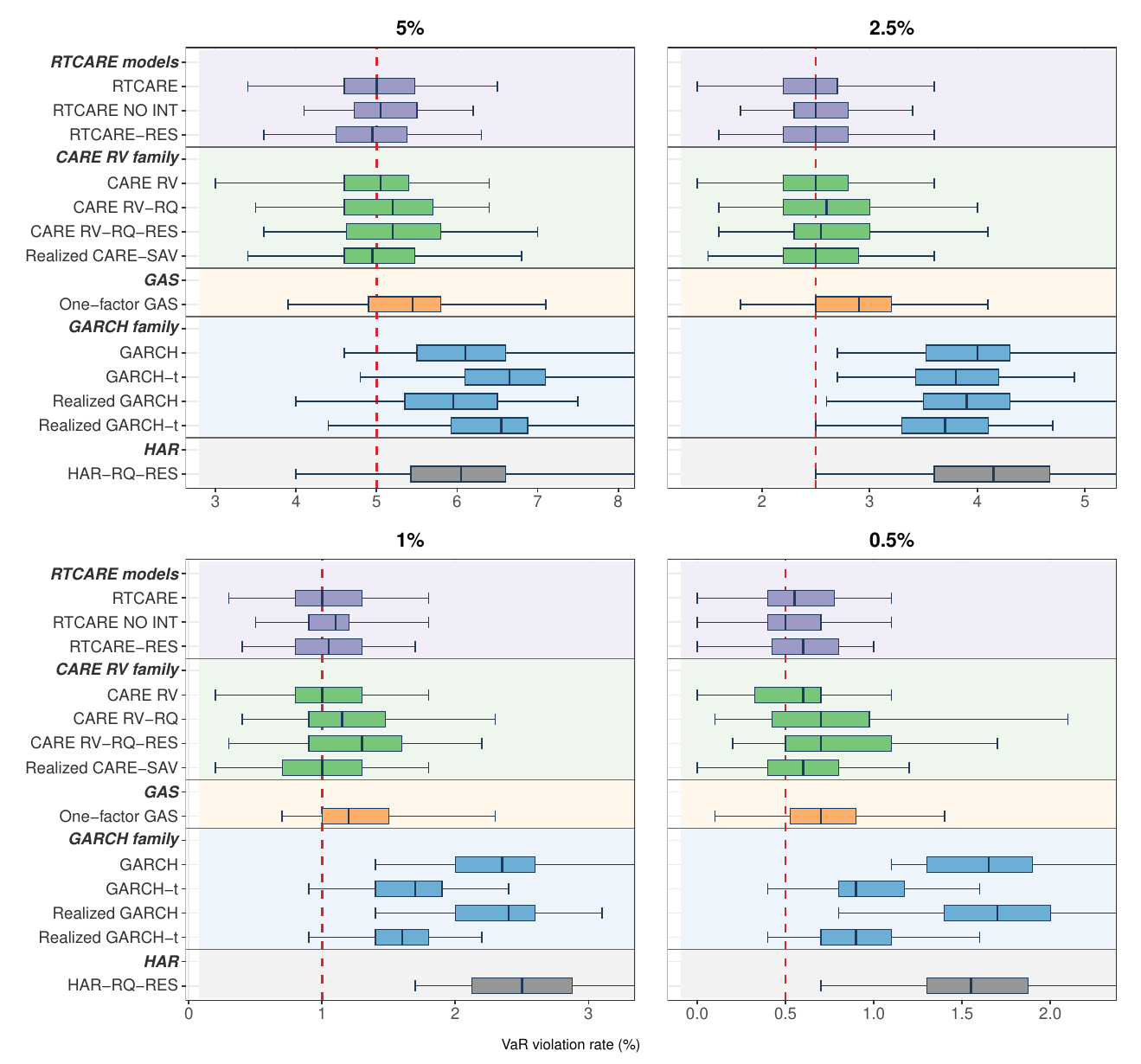}}
      \caption{Boxplots of VaR violation rates across the 50 replications. The red horizontal line marks the nominal violation rate for each tail level.}
      \label{fig:numerical-violation-rate}
      \vspace{7mm}
    \end{figure}

    \noindent
    In summary, Tables~\ref{tab:sim-gjr-fz0}--\ref{tab:sim-gjr-violation}
    show that the RTCARE models perform competitively across the three criteria. They have low FZ0 losses, relatively small $D$ values at all quantile levels, and violation rates close to their nominal values. The generally modest standard deviations across replications support the proposed RTCARE construction as an accurate and stable approach to joint VaR--ES forecasting.\\
    \indent
    Figures~\ref{fig:numerical-fz0}, \ref{fig:numerical-d}, and
    \ref{fig:numerical-violation-rate} present the FZ0 loss, $D$ diagnostic,
    and violation-rate boxplots based on the 50 replication results.  Moreover,
    RMSE distributions for VaR and ES are reported in the Supplementary Material,
    Section~D.
    Figure~\ref{fig:numerical-fz0} shows the mean FZ0 loss across the 50
    replications.  The RTCARE NO INT box is consistently among the lowest,
    while the CARE RV-RQ and CARE RV-RQ-RES boxes widen in the extreme tail.
    GARCH-t and Realized GARCH-t remain competitive in the center of the
    distributions, but the proposed specifications show a more stable loss profile.
    Figure~\ref{fig:numerical-d} provides the boxplots for the $D$ diagnostics.
    The RTCARE boxes remain relatively compact, whereas the CARE RV-RQ,
    CARE RV-RQ-RES, and HAR-RQ-RES boxes become more dispersed at 1\% and 0.5\%.
    Realized GARCH-t is competitive at 0.5\%, consistent with the corresponding
    table entry.
    Figure~\ref{fig:numerical-violation-rate} compares the violation-rate boxes
    with the red nominal-rate lines.  The RTCARE boxes remain close to their
    respective lines, while the GARCH, Realized GARCH, and HAR-RQ-RES boxes are
    generally above the targets.  The wider boxes of the CARE RV-RQ and
    CARE RV-RQ-RES variants show greater replication sensitivity, especially at
    the 0.5\% level where only a few violations are expected in each replication.
    Overall, the boxplots show that the proposed RTCARE models generally combine
    lower average FZ0 losses and $D$ values with violation rates close to the
    nominal targets. Their relatively narrow boxes and whiskers also show less
    variation across replications than many competing specifications. Thus, the
    proposed class offers favorable average performance together with a stable
    profile across the four tail levels.\\
\clearpage

    \newpage
    \section{Application to IBM Stock Returns}
    \label{sec:empirical}
    \subsection{Data}
    The empirical application uses IBM tick data from 4 January 2005 through 31
    December 2024, which is obtained from Tick Data, LLC (https://www.tickdata.com). We calculate daily returns, RQ, and RES from these tick data,
    giving 5013 trading days. 
    Time-series plots of the corresponding daily return, RQ, and RES series
    are provided in the Supplementary Material, Section~E.
    We evaluated one-,
    three-, and five-minute sampling intervals for RQ and RES. Among these
    choices, the three-minute interval produced the fewest outliers and the
    most stable realized measures, and is therefore used throughout the
    empirical analysis.

    \subsection{Estimation Results}
    The proposed RTCARE model\footnotemark is first applied to the full IBM sample
    described above. This full-sample estimation provides parameter estimates
    and posterior summaries before the rolling out-of-sample forecasts are
    considered. We use the four quantile levels
    $\alpha \in \{5\%, 2.5\%, 1\%, 0.5\%\}$. The prior distributions used
    for the MCMC estimation are those specified in Section~\ref{sec:bayes}.
    The resulting posterior summaries are reported in Table~\ref{tab:empirical-param-summary}.

\begin{table}[htbp]
    \centering
    \captionsetup{
      font=tablecaption,
      skip=3pt,
      width=0.85\textwidth
    }
    \caption{Posterior summary statistics for RTCARE (in-sample). The table reports the posterior mean, standard deviation, and 95\% credible interval (2.5\% and 97.5\% quantiles) for $\alpha=5\%, 2.5\%, 1\%, 0.5\%$.}
    \label{tab:empirical-param-summary}
    \fontsize{10}{11}\selectfont
    \setlength{\tabcolsep}{3pt}
    \renewcommand{\arraystretch}{0.88}
    
    \setlength{\tabcolsep}{3pt}
    \begin{tabular*}{0.85\textwidth}{@{\extracolsep{\fill}}lrrrrrrrr}
    \toprule
    & \multicolumn{4}{c}{$\alpha=5\%$} & \multicolumn{4}{c}{$\alpha=2.5\%$} \\
    \cmidrule(lr){2-5} \cmidrule(lr){6-9}
    Parameter & Mean & SD & 2.5\% & 97.5\% & Mean & SD & 2.5\% & 97.5\% \\
    \midrule
    $\beta_0$ & -0.273 & 0.033 & -0.336 & -0.211 & -0.377 & 0.036 & -0.459 & -0.305 \\
    $\beta_1$ & 0.525 & 0.015 & 0.496 & 0.554 & 0.604 & 0.015 & 0.574 & 0.633 \\
    $\beta_2$ & 0.377 & 0.020 & 0.338 & 0.417 & 0.307 & 0.017 & 0.274 & 0.341 \\
    $\alpha_q$ & 0.501 & 0.106 & 0.299 & 0.717 & 0.883 & 0.144 & 0.613 & 1.171 \\
    $\beta_q$ & 1.143 & 0.049 & 1.051 & 1.249 & 1.157 & 0.051 & 1.060 & 1.258 \\
    $\sigma^2_r$ & 0.709 & 0.014 & 0.682 & 0.737 & 0.505 & 0.010 & 0.486 & 0.526 \\
    $\sigma^2_u$ & 0.318 & 0.006 & 0.305 & 0.330 & 0.621 & 0.012 & 0.597 & 0.646 \\
    \bottomrule
    \end{tabular*}
    
    \vspace{0.5em}
    
    \setlength{\tabcolsep}{3pt}
    \begin{tabular*}{0.85\textwidth}{@{\extracolsep{\fill}}lrrrrrrrr}
    \toprule
    & \multicolumn{4}{c}{$\alpha=1\%$} & \multicolumn{4}{c}{$\alpha=0.5\%$} \\
    \cmidrule(lr){2-5} \cmidrule(lr){6-9}
    Parameter & Mean & SD & 2.5\% & 97.5\% & Mean & SD & 2.5\% & 97.5\% \\
    \midrule
    $\beta_0$ & -0.426 & 0.032 & -0.490 & -0.369 & -0.442 & 0.032 & -0.507 & -0.382 \\
    $\beta_1$ & 0.762 & 0.011 & 0.741 & 0.783 & 0.846 & 0.010 & 0.825 & 0.864 \\
    $\beta_2$ & 0.160 & 0.011 & 0.140 & 0.183 & 0.097 & 0.007 & 0.084 & 0.111 \\
    $\alpha_q$ & 1.967 & 0.261 & 1.471 & 2.510 & 2.247 & 0.322 & 1.615 & 2.903 \\
    $\beta_q$ & 1.287 & 0.070 & 1.157 & 1.433 & 1.115 & 0.064 & 0.991 & 1.250 \\
    $\sigma^2_r$ & 0.324 & 0.007 & 0.312 & 0.337 & 0.184 & 0.004 & 0.177 & 0.191 \\
    $\sigma^2_u$ & 2.002 & 0.040 & 1.926 & 2.082 & 15.300 & 0.310 & 14.702 & 15.924 \\
    \bottomrule
    \end{tabular*}
\vspace{3mm}
\end{table}

\footnotetext{RTCARE is defined by $y_t = \mu_t+\varepsilon_t$ ($\varepsilon_t\sim\text{AN}(0,\sigma_r^2,\tau)$), $\mu_t = \beta_0+\beta_1\mu_{t-1}+\beta_2 RQ_{t-1}$, and $RQ_t = \alpha_q+\beta_q\mu_t+u_t$ ($u_t\sim\mathcal{N}(0,\sigma_u^2)$).}

    The posterior means indicate that the coefficient on the lagged expectile, $\beta_1$,
    rises from 0.525 at $\alpha=5\%$ to 0.846 at $\alpha=0.5\%$, while
    the loading on lagged RQ, $\beta_2$, declines from 0.377 to 0.097.
    The measurement
    slope $\beta_q$ remains positive and is close to one, ranging between 1.115
    and 1.287 across the four levels.  The return pseudo-likelihood scale parameter, $\sigma_r^2$, declines
    from 0.709 to 0.184, whereas the measurement-error variance, $\sigma_u^2$, increases
    from 0.318 to 15.300, indicating substantially greater dispersion in RQ at the most extreme level.
    In the RTCARE NO INT model, the omission of the intercept from the measurement equation causes the other parameters to adjust during estimation, and the resulting level of $\mu_t$ is similar to that of the other RTCARE variants.
    \subsection{Comparison of In-Sample Predictive Performance}
    \label{sec:in-sample-forecasting-comparison}
 
    The full-sample estimates describe the fitted tail dynamics. This subsection compares the competing specifications using their in-sample tail-risk estimates and evaluation criteria. Out-of-sample predictive performance is discussed in Section~\ref{sec:empirical-oos}. \\
    \noindent
    {\it FZ0 loss, $D$ diagnostics, and violation rates.}
    Using the same four quantile levels as in the full-sample estimation above, we evaluate the dynamic model specifications considered in the numerical example. For the in-sample comparison, the HAR-RQ-RES benchmark is replaced by the contemporaneous RQ-RES benchmark.  
    Tables~\ref{tab:empirical-is-fz0} and~\ref{tab:empirical-is-d} report the average FZ0 loss and $D$ diagnostics for each of the four quantile levels, while Table~\ref{tab:empirical-is-violation} reports violation rates against their nominal rates. 
    \vspace{3mm}

    \begin{table}[htbp]
    \centering
    \captionsetup{font=tablecaption,skip=3pt,width=0.52\textwidth}
    \caption{Mean FZ0 loss for all quantile levels (in-sample). * indicates the smallest value.}
    \label{tab:empirical-is-fz0}
    \fontsize{10}{11}\selectfont
    \setlength{\tabcolsep}{7pt}
    \renewcommand{\arraystretch}{0.88}
    \begin{tabular}{@{}lrrrr@{}}
    \toprule
    Model & $5\%$ & $2.5\%$ & $1\%$ & $0.5\%$ \\
    \midrule
    RTCARE & 1.072 & 1.333 & 1.683 & 1.915 \\
    RTCARE NO INT & 1.065 & 1.319 & 1.673 & 1.909 \\
    RTCARE-RES & 1.072 & 1.333 & 1.682 & 1.919 \\
\midrule
    CARE RV & 1.137 & 1.400 & 1.718 & 1.940 \\
    CARE RV-RQ & 1.070 & 1.324 & 1.680 & 1.906 \\
    CARE RV-RQ-RES & 1.070 & 1.324 & 1.673 & 1.913 \\
    Realized CARE-SAV & 1.140 & 1.395 & 1.711 & 1.933 \\
\midrule
    One-factor GAS & 1.136 & 1.422 & 1.768 & 1.970 \\
\midrule
    GARCH & 1.153 & 1.434 & 1.896 & 2.408 \\
    GARCH-t & 1.135 & 1.405 & 1.774 & 2.061 \\
    Realized GARCH & 1.111 & 1.378 & 1.823 & 2.315 \\
    Realized GARCH-t & 1.095 & 1.349 & 1.694 & 1.959 \\
\midrule
    RQ-RES benchmark & *0.743 & *0.929 & *1.178 & *1.422 \\
    \bottomrule
    \end{tabular}
    \end{table}
    \vspace{3mm}
    \begin{table}[htbp]
    \centering
    \captionsetup{font=tablecaption,skip=3pt,width=0.52\textwidth}
    \centering
    \caption{$D$ diagnostics at all four quantile levels (in-sample).  * indicates the smallest value.}
    \label{tab:empirical-is-d}
    \fontsize{10}{11}\selectfont
    \setlength{\tabcolsep}{7pt}
    \renewcommand{\arraystretch}{0.88}
    \begin{tabular}{@{}lrrrr@{}}
    \toprule
    Model & $5\%$ & $2.5\%$ & $1\%$ & $0.5\%$ \\
    \midrule
    RTCARE & 0.044 & 0.046 & 0.046 & 0.080 \\
    RTCARE NO INT & 0.152 & 0.188 & 0.274 & *0.037 \\
    RTCARE-RES & *0.024 & 0.059 & 0.066 & 0.041 \\
\midrule
    CARE RV & 0.048 & *0.026 & *0.027 & 0.090 \\
    CARE RV-RQ & 0.049 & 0.027 & 0.085 & 0.066 \\
    CARE RV-RQ-RES & 0.038 & 0.060 & 0.050 & 0.041 \\
    Realized CARE-SAV & 0.276 & 0.280 & 0.195 & 0.243 \\
\midrule
    One-factor GAS & 0.145 & 0.119 & 0.067 & 0.117 \\
\midrule
    GARCH & 0.557 & 0.963 & 1.608 & 2.269 \\
    GARCH-t & 0.191 & 0.338 & 0.752 & 0.950 \\
    Realized GARCH & 0.523 & 0.925 & 1.662 & 2.264 \\
    Realized GARCH-t & 0.150 & 0.367 & 0.737 & 1.072 \\
\midrule
    RQ-RES benchmark & 0.114 & 0.244 & 0.500 & 0.845 \\
    \bottomrule
    \end{tabular}
    \end{table}
    \begin{table}[htbp]
    \centering
    \captionsetup{font=tablecaption,skip=3pt,width=0.52\textwidth}
    \centering
    \caption{Violation rates (in percent) at all four quantile levels (in-sample). * indicates the value closest to the nominal violation rate in each column.}
    \label{tab:empirical-is-violation}
    \fontsize{10}{11}\selectfont
    \setlength{\tabcolsep}{7pt}
    \renewcommand{\arraystretch}{0.88}
    \begin{tabular}{@{}lrrrr@{}}
    \toprule
    Model & $5\%$ & $2.5\%$ & $1\%$ & $0.5\%$ \\
    \midrule
    RTCARE & 5.067 & 2.474 & *0.997 & *0.499 \\
    RTCARE NO INT & *5.007 & *2.494 & 1.017 & *0.499 \\
    RTCARE-RES & 4.867 & 2.513 & 1.017 & *0.499 \\
\midrule
    CARE RV & 4.987 & 2.474 & *0.997 & 0.519 \\
    CARE RV-RQ & 5.107 & 2.454 & *0.997 & *0.499 \\
    CARE RV-RQ-RES & 5.047 & 2.573 & *0.997 & 0.539 \\
    Realized CARE-SAV & 5.047 & *2.494 & *0.997 & 0.539 \\
\midrule
    One-factor GAS & 5.725 & 2.753 & 1.057 & 0.479 \\
\midrule
    GARCH & 3.730 & 2.334 & 1.576 & 1.177 \\
    GARCH-t & 4.628 & 2.454 & 1.117 & 0.738 \\
    Realized GARCH & 3.850 & 2.354 & 1.416 & 1.157 \\
    Realized GARCH-t & 4.907 & 2.394 & 1.117 & 0.658 \\
\midrule
    RQ-RES benchmark & 4.947 & 2.833 & 1.237 & 0.738 \\
    \bottomrule
    \end{tabular}
    \vspace{3mm}
    \end{table}
   \indent
   For the FZ0 loss in Table \ref{tab:empirical-is-fz0}, the RQ-RES benchmark based on same-day realized tail-risk measures has the smallest loss at every level, ranging from 0.743 at 5\% to 1.422 at 0.5\%. Among the dynamic specifications, RTCARE NO INT has the smallest loss at 5\% and 2.5\%, ties with CARE RV-RQ-RES at 1\% at the displayed precision, and is surpassed by CARE RV-RQ at 0.5\%. The other RTCARE and CARE RV-RQ variants remain close to these values. The GARCH and Realized GARCH specifications become less competitive as the tail becomes more extreme.

    \indent 
    For $D$ diagnostics in Table \ref{tab:empirical-is-d}, 
    the RTCARE and CARE RV-RQ specifications generally have small $D$ values, whereas the $D$ values for the normal GARCH and Realized GARCH specifications increase markedly at the more extreme levels. RTCARE-RES has the smallest $D$ value at 5\%, CARE RV at 2.5\% and 1\%, and RTCARE NO INT at 0.5\%. CARE RV-RQ-RES and the other RTCARE variants remain competitive on this diagnostic.

    Finally, the violation rates in Table~\ref{tab:empirical-is-violation} show that the RTCARE variants stay close to their nominal targets across all four levels. The GARCH and Realized GARCH models have violation rates increasingly above the nominal levels as $\alpha$ decreases, whereas their $t$ versions reduce but do not remove this deviation. The same-day RQ-RES benchmark is close at 5\% but becomes less well calibrated at the more extreme levels.

    Overall, the three tables show that the proposed RTCARE variants have small FZ0 losses and $D$ diagnostics, and that their violation rates are close to the nominal targets across tail levels. The lower FZ0 loss of the benchmark based on same-day RQ and RES is an in-sample advantage that does not carry over to one-day-ahead forecasting. The corresponding rank plots are reported in the Supplementary Material, Section~F.

    \begin{table}[htbp]
    \centering
    \captionsetup{font=tablecaption,skip=3pt,width=0.90\textwidth}
    \caption{Average ranks for $\alpha \in \{5\%, 2.5\%, 1\%, 0.5\%\}$ (in-sample). Boldface indicates the best rank for each column.}
    \label{tab:is_avg_ranking}
    \fontsize{10}{11}\selectfont
    \setlength{\tabcolsep}{3pt}
    \renewcommand{\arraystretch}{0.88}
    \begin{tabular*}{0.90\textwidth}{@{\extracolsep{\fill}}lrrrr}
    \toprule
    Model & Avg FZ0 Rank & Avg $D$ Rank & Avg Violation Rank & Overall Avg Rank \\
    \midrule
    RTCARE & 5.25 & 3.25 & 3.00 & 3.83 \\
    RTCARE NO INT & 2.25 & 6.25 & \textbf{2.25} & \textbf{3.58} \\
    RTCARE-RES & 5.75 & \textbf{3.00} & 4.75 & 4.50 \\
    \midrule
    CARE RV & 9.50 & \textbf{3.00} & 3.00 & 5.17 \\
    CARE RV-RQ & 3.00 & 4.25 & 4.00 & 3.75 \\
    CARE RV-RQ-RES & 3.75 & \textbf{3.00} & 4.75 & 3.83 \\
    Realized CARE-SAV & 9.00 & 8.75 & 3.00 & 6.92 \\
    \midrule
    One-factor GAS & 10.50 & 6.25 & 9.25 & 8.67 \\
    \midrule
    GARCH & 13.00 & 12.75 & 12.50 & 12.75 \\
    GARCH-t & 10.50 & 10.25 & 8.75 & 9.83 \\
    Realized GARCH & 10.00 & 12.25 & 11.50 & 11.25 \\
    Realized GARCH-t & 7.50 & 10.00 & 8.50 & 8.67 \\
    \midrule
    RQ-RES benchmark & \textbf{1.00} & 8.00 & 9.75 & 6.25 \\
    \bottomrule
    \end{tabular*}
    \end{table}
    \renewcommand{\arraystretch}{1}
    \spacingset{1}

    Table~\ref{tab:is_avg_ranking} reports ranks averaged over the four tail levels and, overall, over the three criteria. Ranking conventions are given in the Supplementary Material, Section~F. RTCARE NO INT has the best overall in-sample rank, followed by CARE RV-RQ, while RTCARE and CARE RV-RQ-RES tie at the next position. The RQ-RES benchmark ranks first on FZ0 because it uses contemporaneous realized tail measures, but its $D$ and violation-rate ranks are weaker.

    \noindent
    {\it Regression-based ES backtesting}. To assess whether the in-sample ES estimates are correctly specified, we use the three regression-based tests proposed by \citet{BayerDimitriadis22}: (1) the strict test, (2) the auxiliary test and (3) the intercept test. 
    The tests use the joint VaR--ES regression framework described in the Supplementary Material, Section~G.
    In Table~\ref{tab:empirical-is-backtest-mark-only}, we report whether we fail to reject the null hypothesis, $H_0:$ the in-sample ES estimates are correctly specified, at the 5\% significance level (two-sided tests are considered in the intercept tests). 
      
    \begin{table}[htbp]
    \providecommand{\passmark}{\ensuremath{\bullet}}
    \centering
    \captionsetup{
      font=tablecaption,
      skip=3pt,
      width=0.78\textwidth
    }
    \caption{Results of the ES backtests proposed by \citet{BayerDimitriadis22} for $\alpha=5\%, 2.5\%, 1\%, 0.5\%$ (in-sample). The check mark (\passmark) indicates that we fail to reject the null hypothesis, $H_0:$ the in-sample ES estimates are correctly specified, at the 5\% significance level.}
    \label{tab:empirical-is-backtest-mark-only}
    \fontsize{10}{11}\selectfont
    \setlength{\tabcolsep}{3pt}
    \renewcommand{\arraystretch}{0.88}
    \setlength{\tabcolsep}{3pt}

    \begin{tabular}{@{}lrrrrrr@{}}
    \toprule
    Model & \multicolumn{3}{c}{$\alpha=5\%$} & \multicolumn{3}{c}{$\alpha=2.5\%$} \\
    \cmidrule(lr){2-4} \cmidrule(lr){5-7}
    & Strict & Auxiliary & Intercept & Strict & Auxiliary & Intercept \\
    \midrule
    RTCARE & \passmark & \passmark & \passmark & \passmark & \passmark & \passmark \\
    RTCARE NO INT & \passmark & \passmark & \passmark & \passmark & \passmark & \passmark \\
    RTCARE-RES & \passmark & \passmark & \passmark & \passmark & \passmark & \passmark \\
\midrule
    CARE RV & \passmark & \passmark & \passmark & \passmark & \passmark & \passmark \\
    CARE RV-RQ & \passmark & \passmark & \passmark & \passmark & \passmark & \passmark \\
    CARE RV-RQ-RES & \passmark & \passmark & \passmark & \passmark & \passmark & \passmark \\
    Realized CARE-SAV & -- & -- & -- & \passmark & \passmark & \passmark \\
\midrule
    One-factor GAS & \passmark & \passmark & \passmark & \passmark & \passmark & \passmark \\
\midrule
    GARCH & -- & -- & -- & -- & -- & -- \\
    GARCH-t & \passmark & \passmark & \passmark & \passmark & \passmark & \passmark \\
    Realized GARCH & -- & -- & -- & -- & -- & -- \\
    Realized GARCH-t & \passmark & \passmark & \passmark & \passmark & \passmark & \passmark \\
\midrule
    RQ-RES benchmark & -- & \passmark & \passmark & -- & \passmark & -- \\
    \bottomrule
    \end{tabular}
    \vspace{1.5mm} \\
    
    \begin{tabular}{@{}lrrrrrr@{}}
    \toprule
    Model & \multicolumn{3}{c}{$\alpha=1\%$} & \multicolumn{3}{c}{$\alpha=0.5\%$} \\
    \cmidrule(lr){2-4} \cmidrule(lr){5-7}
    & Strict & Auxiliary & Intercept & Strict & Auxiliary & Intercept \\
    \midrule
    RTCARE & \passmark & \passmark & \passmark & \passmark & \passmark & \passmark \\
    RTCARE NO INT & \passmark & \passmark & \passmark & \passmark & \passmark & \passmark \\
    RTCARE-RES & \passmark & \passmark & \passmark & \passmark & \passmark & \passmark \\
\midrule
    CARE RV & \passmark & \passmark & \passmark & \passmark & \passmark & \passmark \\
    CARE RV-RQ & \passmark & \passmark & \passmark & \passmark & \passmark & \passmark \\
    CARE RV-RQ-RES & \passmark & \passmark & \passmark & \passmark & \passmark & \passmark \\
    Realized CARE-SAV & \passmark & \passmark & \passmark & \passmark & \passmark & \passmark \\
\midrule
    One-factor GAS & \passmark & \passmark & \passmark & \passmark & \passmark & \passmark \\
\midrule
    GARCH & -- & -- & -- & -- & -- & -- \\
    GARCH-t & -- & -- & \passmark & -- & -- & \passmark \\
    Realized GARCH & -- & -- & -- & -- & -- & -- \\
    Realized GARCH-t & \passmark & \passmark & -- & -- & -- & -- \\
\midrule
    RQ-RES benchmark & -- & \passmark & \passmark & \passmark & \passmark & \passmark \\
    \bottomrule
    \end{tabular}
    \vspace{5mm}
\end{table}
    \renewcommand{\arraystretch}{1}
    \spacingset{1}
    \noindent
    The check marks show no rejection for the RTCARE specifications, the CARE RV variants, or the GAS model at any of the four levels. The normal GARCH and Realized GARCH specifications are rejected by all three procedures. The GARCH-t and Realized GARCH-t specifications are not rejected at the less extreme levels but are rejected by some tests at 1\% and 0.5\%, while the Realized CARE-SAV model is rejected at 5\% but passes at 2.5\%, 1\%, and 0.5\%. The same-day RQ-RES benchmark passes the auxiliary and intercept tests at most levels, but the null hypothesis of the strict test is rejected except at 0.5\%.

   \subsection{Comparison of Out-of-Sample Predictive Performance}
    \label{sec:empirical-oos}
    The empirical analysis focuses primarily on out-of-sample forecasting because it evaluates predictions for observations that were not used for estimation. 
    We first compare the forecast diagnostics, then summarize the
    model rankings, assess calibration with the ES backtests, and finally report the model confidence sets.
    Each model is re-estimated in a rolling window of 1500 days and
    produces 3513 one-day-ahead forecasts. The reported evaluation statistics are based on predictions formed using only information available before the forecast date.
    \\
    \noindent
    {\it FZ0 loss, $D$ diagnostics, and violation rates.}
     Tables~\ref{tab:empirical-oos-fz0} and~\ref{tab:empirical-oos-d}
    report the FZ0 and $D$ diagnostics across the four quantile levels, while Table~\ref{tab:empirical-oos-violation} reports violation rates with their nominal target rates.

\begin{table}[htbp]
\centering
    \captionsetup{
      font=tablecaption,
      skip=3pt,
  width=0.52\textwidth
    }
    \caption{Mean FZ0 loss for all quantile levels (out-of-sample). * indicates the smallest value.}
    \label{tab:empirical-oos-fz0}
    \fontsize{10}{11}\selectfont
\setlength{\tabcolsep}{7pt}
    \renewcommand{\arraystretch}{0.88}
    \begin{tabular}{@{}lrrrr@{}}
    \toprule
    Model & $5\%$ & $2.5\%$ & $1\%$ & $0.5\%$ \\
    \midrule
    RTCARE & 1.100 & 1.379 & 1.781 & 2.057 \\
    RTCARE NO INT & *1.097 & *1.372 & *1.769 & 2.057 \\
    RTCARE-RES & 1.101 & 1.382 & 1.782 & *2.054 \\
\midrule
    CARE RV & 1.184 & 1.482 & 1.833 & 2.075 \\
    CARE RV-RQ & 1.108 & 1.393 & 1.867 & 2.274 \\
    CARE RV-RQ-RES & 1.124 & 1.411 & 1.848 & 2.150 \\
    Realized CARE-SAV & 1.185 & 1.474 & 1.827 & 2.058 \\
\midrule
    One-factor GAS & 1.211 & 1.526 & 1.925 & 2.302 \\
\midrule
    GARCH & 1.216 & 1.544 & 2.136 & 2.821 \\
    GARCH-t & 1.188 & 1.485 & 1.892 & 2.218 \\
    Realized GARCH & 1.182 & 1.504 & 2.065 & 2.734 \\
    Realized GARCH-t & 1.163 & 1.450 & 1.840 & 2.147 \\
\midrule
    HAR-RQ-RES & 1.195 & 1.593 & 2.299 & 2.884 \\
    \bottomrule
    \end{tabular}

    \end{table}
    \vspace{3mm}
    \begin{table}[htbp]
    \centering
    \captionsetup{
      font=tablecaption,
      skip=3pt,
  width=0.52\textwidth
    }
    \caption{$D$ diagnostics at all four quantile levels (out-of-sample).  * indicates the smallest value.}
    \label{tab:empirical-oos-d}
    \fontsize{10}{11}\selectfont
\setlength{\tabcolsep}{7pt}
    \renewcommand{\arraystretch}{0.88}
    \begin{tabular}{@{}lrrrr@{}}
    \toprule
    Model & $5\%$ & $2.5\%$ & $1\%$ & $0.5\%$ \\
    \midrule
    RTCARE & 0.110 & 0.214 & 0.500 & *0.853 \\
    RTCARE NO INT & 0.247 & 0.326 & 0.606 & 0.931 \\
    RTCARE-RES & 0.125 & 0.207 & *0.484 & 0.962 \\
\midrule
    CARE RV & 0.163 & 0.263 & 0.493 & 1.001 \\
    CARE RV-RQ & 0.212 & 0.416 & 0.753 & 1.224 \\
    CARE RV-RQ-RES & 0.228 & 0.389 & 0.693 & 1.428 \\
    Realized CARE-SAV & *0.106 & *0.165 & 0.552 & 0.866 \\
\midrule
    One-factor GAS & 0.146 & 0.185 & 0.641 & 1.339 \\
\midrule
    GARCH & 0.765 & 1.237 & 1.990 & 2.647 \\
    GARCH-t & 0.368 & 0.602 & 0.999 & 1.237 \\
    Realized GARCH & 0.786 & 1.190 & 2.055 & 2.705 \\
    Realized GARCH-t & 0.450 & 0.654 & 1.121 & 1.121 \\
\midrule
    HAR-RQ-RES & 0.675 & 1.054 & 1.680 & 2.251 \\
    \bottomrule
    \end{tabular}
\end{table}

    \begin{table}[htbp]
    \centering
    \captionsetup{
      font=tablecaption,
      skip=3pt,
  width=0.52\textwidth
    }
    \caption{Violation rates (in percent) at all four quantile levels (out-of-sample). * indicates the value closest to the nominal violation rate in each column.}
    \label{tab:empirical-oos-violation}
    \fontsize{10}{11}\selectfont
\setlength{\tabcolsep}{7pt}
    \renewcommand{\arraystretch}{0.88}
    \begin{tabular}{@{}lrrrr@{}}
    \toprule
    Model & $5\%$ & $2.5\%$ & $1\%$ & $0.5\%$ \\
    \midrule
    RTCARE & 4.782 & 2.420 & *1.082 & 0.626 \\
    RTCARE NO INT & *5.095 & 2.676 & 1.224 & *0.598 \\
    RTCARE-RES & 4.668 & 2.448 & *1.082 & *0.598 \\
\midrule
    CARE RV & 4.668 & *2.505 & 1.167 & 0.626 \\
    CARE RV-RQ & 4.782 & 2.420 & 1.167 & 0.740 \\
    CARE RV-RQ-RES & 4.611 & *2.505 & 1.196 & 0.712 \\
    Realized CARE-SAV & 4.839 & 2.676 & 1.110 & 0.626 \\
\midrule
    One-factor GAS & 5.465 & 2.932 & 1.196 & 0.655 \\
\midrule
    GARCH & 3.985 & 2.647 & 1.879 & 1.423 \\
    GARCH-t & 4.811 & 2.619 & 1.309 & 0.911 \\
    Realized GARCH & 4.327 & 3.046 & 1.850 & 1.423 \\
    Realized GARCH-t & 5.124 & 3.017 & 1.423 & 1.082 \\
\midrule
    HAR-RQ-RES & 7.743 & 4.953 & 2.960 & 1.765 \\
    \bottomrule
    \end{tabular}
    \vspace{3mm}
    \end{table}
   \indent
   Tables~\ref{tab:empirical-oos-fz0} and~\ref{tab:empirical-oos-d} show that the proposed RTCARE class performs well in the out-of-sample comparison. The three variants remain in the leading FZ0 group at every tail level, and the differences within the class are small.  RTCARE NO INT has the lowest FZ0 loss at 5\%, 2.5\%, and 1\%, while RTCARE-RES is lowest at 0.5\%.  
   Forecasts from the GARCH, Realized GARCH, and HAR-RQ-RES models deteriorate markedly as the tail becomes more extreme.
   The $D$ columns give further information about this result. The RTCARE variants remain in the leading group, especially at the 1\% and 0.5\% levels, while the GARCH family and HAR-RQ-RES model generally have larger $D$ values.  Realized CARE-SAV is also competitive on $D$ at several levels, whereas the CARE RV variants are less consistent across the three diagnostics.

   In Table~\ref{tab:empirical-oos-violation}, the violation rates for the three RTCARE models stay close to the nominal target rate at all four levels, including the extreme levels where those of the GARCH, Realized GARCH, and HAR-RQ-RES models substantially exceed the nominal levels. This coverage pattern complements the FZ0 and $D$ results. The three RTCARE models combine accurate coverage with competitive performance across all four quantile levels.
    The average ranks are summarized in Table~\ref{tab:oos_avg_ranking} (see 
    the corresponding rank plots reported in the Supplementary Material, Section~F). 
    It places RTCARE first overall, followed by RTCARE-RES and RTCARE NO INT. This ordering is driven mainly by consistently low FZ0 losses and competitive $D$ diagnostics. Realized CARE-SAV remains competitive on $D$ values and violation rates, while the GARCH and HAR-RQ-RES models rank lower after the three criteria are combined.
    \vspace{10mm}
    \begin{table}[htbp]
    \centering
    \captionsetup{font=tablecaption,skip=3pt,width=0.90\textwidth}
    \caption{Average ranks for $\alpha \in \{5\%, 2.5\%, 1\%, 0.5\%\}$ (out-of-sample). Boldface indicates the best rank for each column.}
    \label{tab:oos_avg_ranking}
    \fontsize{10}{11}\selectfont
    \setlength{\tabcolsep}{3pt}
    \renewcommand{\arraystretch}{0.88}
    \begin{tabular*}{0.90\textwidth}{@{\extracolsep{\fill}}lrrrr}
    \toprule
    Model & Avg FZ0 Rank & Avg $D$ Rank & Avg Violation Rank & Overall Avg Rank \\
    \midrule
    RTCARE & 2.00 & 2.50 & 3.25 & \textbf{2.58} \\
    RTCARE NO INT & \textbf{1.50} & 5.50 & 4.50 & 3.83 \\
    RTCARE-RES & 2.50 & 2.75 & \textbf{3.00} & 2.75 \\
    \midrule
    CARE RV & 6.50 & 4.25 & 3.75 & 4.83 \\
    CARE RV-RQ & 6.25 & 7.25 & 5.25 & 6.25 \\
    CARE RV-RQ-RES & 6.00 & 7.75 & 5.75 & 6.50 \\
    Realized CARE-SAV & 6.00 & \textbf{2.00} & 4.25 & 4.08 \\
    \midrule
    One-factor GAS & 10.75 & 5.25 & 8.00 & 8.00 \\
    \midrule
    GARCH & 12.25 & 12.25 & 10.50 & 11.67 \\
    GARCH-t & 9.00 & 8.75 & 7.00 & 8.25 \\
    Realized GARCH & 9.75 & 12.75 & 11.25 & 11.25 \\
    Realized GARCH-t & 6.00 & 9.00 & 8.25 & 7.75 \\
    \midrule
    HAR-RQ-RES & 12.50 & 11.00 & 13.00 & 12.17 \\
    \bottomrule
    \end{tabular*}
    \vspace{5mm}
    \end{table}
    \renewcommand{\arraystretch}{1}
    \spacingset{1}

    \noindent
    {\it Regression-based ES backtesting}. 
    The three regression-based backtesting procedures proposed by \citet{BayerDimitriadis22} are also applied to the rolling one-day-ahead forecasts as in Section \ref{sec:in-sample-forecasting-comparison}. Their results are reported in Table~\ref{tab:empirical-oos-backtest-mark-only}. A check mark indicates that the corresponding calibration null is not rejected at the 5\% significance level.
\begin{table}[htbp]
    \providecommand{\passmark}{\ensuremath{\bullet}}
    \centering
     \captionsetup{
       font=tablecaption,
       skip=3pt,
       width=0.78\textwidth
     }
     \caption{Results of the ES backtests proposed by \citet{BayerDimitriadis22} for $\alpha=5\%, 2.5\%, 1\%, 0.5\%$ (out-of-sample). The check mark (\passmark) indicates that we fail to reject the null hypothesis, $H_0:$ ES forecasts are correctly specified, at the 5\% significance level.}
     \label{tab:empirical-oos-backtest-mark-only}
     \fontsize{10}{11}\selectfont
\setlength{\tabcolsep}{3pt}
     \renewcommand{\arraystretch}{0.88}
\setlength{\tabcolsep}{3pt}

    \begin{tabular}{@{}lrrrrrr@{}}
    \toprule
    Model & \multicolumn{3}{c}{$\alpha=5\%$} & \multicolumn{3}{c}{$\alpha=2.5\%$} \\
    \cmidrule(lr){2-4} \cmidrule(lr){5-7}
    & Strict & Auxiliary & Intercept & Strict & Auxiliary & Intercept \\
    \midrule
    RTCARE & \passmark & \passmark & \passmark & \passmark & \passmark & \passmark \\
    RTCARE NO INT & \passmark & \passmark & \passmark & \passmark & \passmark & \passmark \\
    RTCARE-RES & \passmark & \passmark & \passmark & \passmark & \passmark & \passmark \\
\midrule
    CARE RV & \passmark & \passmark & \passmark & \passmark & \passmark & \passmark \\
    CARE RV-RQ & \passmark & \passmark & \passmark & \passmark & -- & \passmark \\
    CARE RV-RQ-RES & \passmark & -- & \passmark & -- & -- & \passmark \\
    Realized CARE-SAV & \passmark & \passmark & \passmark & \passmark & \passmark & \passmark \\
\midrule
    One-factor GAS & \passmark & \passmark & \passmark & \passmark & \passmark & \passmark \\
\midrule
    GARCH & -- & -- & -- & -- & -- & -- \\
    GARCH-t & \passmark & \passmark & \passmark & \passmark & -- & \passmark \\
    Realized GARCH & -- & -- & -- & -- & -- & -- \\
    Realized GARCH-t & -- & -- & -- & -- & -- & -- \\
\midrule
    HAR-RQ-RES & -- & -- & -- & -- & -- & -- \\
    \bottomrule
    \end{tabular}
    
    \vspace{2.5mm}
    
    \begin{tabular}{@{}lrrrrrr@{}}
    \toprule
    Model & \multicolumn{3}{c}{$\alpha=1\%$} & \multicolumn{3}{c}{$\alpha=0.5\%$} \\
    \cmidrule(lr){2-4} \cmidrule(lr){5-7}
    & Strict & Auxiliary & Intercept & Strict & Auxiliary & Intercept \\
    \midrule
    RTCARE & \passmark & \passmark & \passmark & \passmark & \passmark & \passmark \\
    RTCARE NO INT & \passmark & \passmark & \passmark & \passmark & \passmark & \passmark \\
    RTCARE-RES & \passmark & \passmark & \passmark & \passmark & \passmark & \passmark \\
\midrule
    CARE RV & \passmark & \passmark & \passmark & \passmark & \passmark & \passmark \\
    CARE RV-RQ & \passmark & \passmark & \passmark & \passmark & \passmark & \passmark \\
    CARE RV-RQ-RES & \passmark & \passmark & \passmark & \passmark & \passmark & \passmark \\
    Realized CARE-SAV & \passmark & \passmark & \passmark & \passmark & \passmark & \passmark \\
\midrule
    One-factor GAS & \passmark & \passmark & \passmark & -- & -- & \passmark \\
\midrule
    GARCH & -- & -- & -- & -- & -- & -- \\
    GARCH-t & -- & -- & -- & -- & -- & -- \\
    Realized GARCH & -- & -- & -- & -- & -- & -- \\
    Realized GARCH-t & -- & -- & -- & -- & -- & \passmark \\
\midrule
    HAR-RQ-RES & -- & -- & -- & -- & -- & -- \\
    \bottomrule
    \end{tabular}
    \vspace{3mm}
\end{table}
    \renewcommand{\arraystretch}{1}
    \spacingset{1}

Many models, including the proposed specifications, are not rejected at most tail levels. In contrast, the HAR-RQ-RES benchmark model, which independently forecasts RQ and RES using HAR dynamics, is rejected by all three tests at every tail level. This contrasts with the same-day RQ-RES benchmark, which has the smallest in-sample FZ0 loss. Its advantage does not carry over when RQ and RES must be forecast out of sample. The backtests therefore reinforce the ranking evidence while also showing that a low score alone is not sufficient for a satisfactory VaR--ES forecast. \\ 

    \noindent
    {\it Model confidence sets}.
    To compare predictive ability across all models while allowing for sampling uncertainty, we use the model confidence set (MCS) procedure of \citet{HansenLundeNason11}. The MCS is the set of models that cannot be statistically distinguished from the best model at a specified confidence level. We apply the range statistic $T_R$ to the common out-of-sample daily FZ0-loss series. $T_R$ is based on the largest studentized pairwise loss
    difference, and the equal-predictive-ability null is evaluated with the bootstrap procedure implemented in the \texttt{MCS} package of \citet{BernardiCatania18}. We use 5,000 bootstrap replications, the package's automatically selected block length, and a minimum block length of three to preserve dependence in the loss series. The algorithm sequentially removes the model most responsible for rejecting equal predictive ability and stops when the remaining set is no longer rejected.
    A model is retained in the 90\% (75\%) MCS when its $p$-value is at least 0.10 (0.25). Both sets are reported in
    Table~\ref{tab:empirical-oos-mcs-90-75-mark-only}. One check mark denotes retention only in the 90\% set, whereas two check marks denote retention in the stricter 75\% set as well.
    \vspace{3mm}

\begin{table}[htbp]
    \providecommand{\mcsmark}{\ensuremath{\circ}}
    \centering
    \captionsetup{
      font=tablecaption,
      skip=3pt,
      width=0.7\textwidth
    }
    \caption{IBM out-of-sample 90\% and 75\% Model Confidence Set (MCS) membership for the $T_R$ statistic. One check mark (\mcsmark) indicates retention in the 90\% MCS only; two check marks (\mcsmark\mcsmark) indicate retention in the 75\% MCS (and therefore also the 90\% MCS).}
    \label{tab:empirical-oos-mcs-90-75-mark-only}
    \begin{tabular}{@{}lrrrr@{}}
    \toprule
    Model & $\alpha=5\%$ & $\alpha=2.5\%$ & $\alpha=1\%$ & $\alpha=0.5\%$ \\
    \midrule
    RTCARE & \mcsmark\mcsmark & \mcsmark\mcsmark & \mcsmark\mcsmark & \mcsmark\mcsmark \\
    RTCARE NO INT & \mcsmark\mcsmark & \mcsmark\mcsmark & \mcsmark\mcsmark & \mcsmark\mcsmark \\
    RTCARE-RES & \mcsmark\mcsmark & \mcsmark\mcsmark & \mcsmark\mcsmark & \mcsmark\mcsmark \\
    CARE RV & -- & -- & \mcsmark\mcsmark & \mcsmark\mcsmark \\
    CARE RV-RQ & \mcsmark\mcsmark & \mcsmark\mcsmark & \mcsmark\mcsmark & \mcsmark\mcsmark \\
    CARE RV-RQ-RES & \mcsmark\mcsmark & \mcsmark & \mcsmark\mcsmark & \mcsmark\mcsmark \\
    Realized CARE-SAV & -- & -- & \mcsmark\mcsmark & \mcsmark\mcsmark \\
    One-factor GAS & -- & -- & \mcsmark & \mcsmark\mcsmark \\
    GARCH & -- & -- & -- & -- \\
    GARCH-t & -- & -- & -- & \mcsmark\mcsmark \\
    Realized GARCH & -- & -- & -- & -- \\
    Realized GARCH-t & -- & -- & \mcsmark\mcsmark & \mcsmark\mcsmark \\
    HAR-RQ-RES & -- & -- & -- & -- \\
    \bottomrule
    \end{tabular}
    \vspace{5mm}
\end{table}
    \renewcommand{\arraystretch}{1}
    \spacingset{1}

    Table~\ref{tab:empirical-oos-mcs-90-75-mark-only} gives a compact comparison of the predictive-ability results. All three proposed RTCARE models are retained in both the 90\% and 75\% sets at every tail probability. 
    The CARE RV variants are retained more selectively. Their membership is strongest at the most extreme tail, while the Realized CARE-SAV and one-factor GAS models are retained mainly at $\alpha=1\%$ and $0.5\%$. 
    The normal GARCH and Realized GARCH specifications are excluded at every level. GARCH-t and Realized GARCH-t are retained only at the most extreme tail levels, with the latter also retained at $\alpha=1\%$. 
    The MCS evidence is consistent with and complements the results of the FZ0 loss, $D$ diagnostics, violation rates and ES backtests reported above.\\
    
\begin{figure}[htbp]
      \centering
      \includegraphics[width=0.95\textwidth]{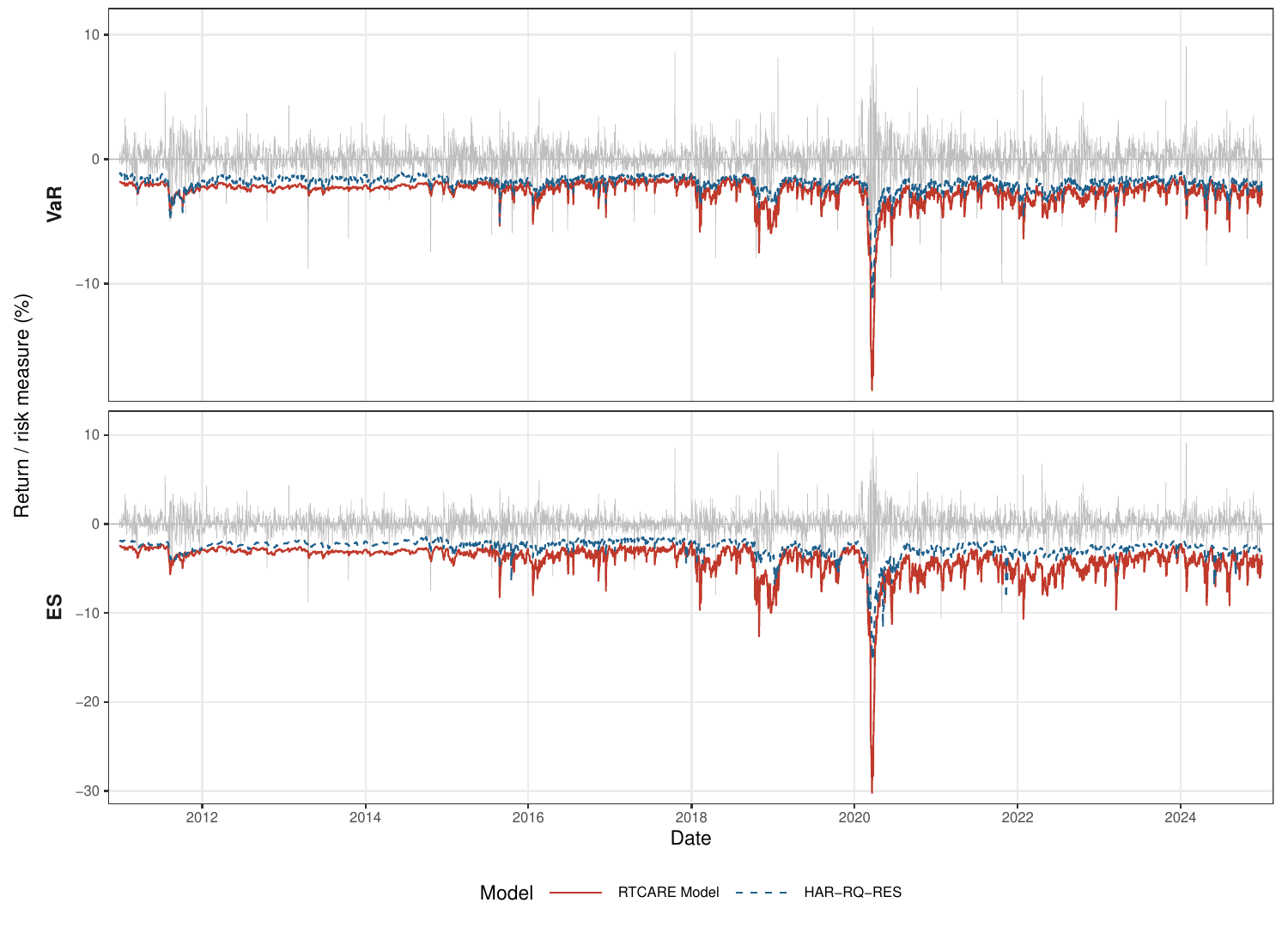}
      \caption{One-day-ahead VaR and ES forecasts at $\alpha=2.5\%$ (out-of-sample). The upper and lower panels show VaR and ES, respectively. Solid gray: daily return series. Solid red: forecasts from RTCARE. Dashed blue: forecasts from HAR-RQ-RES.}
      \label{fig:rcare-rq-forecast}
      \vspace{5mm}
    \end{figure}
    \noindent
    {\it Time-series plot of out-of-sample forecasts}. 
    Finally, Figure~\ref{fig:rcare-rq-forecast} illustrates the rolling forecasts and compares the proposed specification with the
    HAR-RQ-RES benchmark model. Both models produce more negative VaR and ES forecasts during periods of elevated market volatility and return toward their usual levels as conditions calm. ES remains below VaR throughout the period, as required by the ordering of the two risk measures. The HAR-RQ-RES forecasts are visibly closer to zero than the corresponding RTCARE forecasts in both panels. Taken together with the lower loss and stronger diagnostic performance of RTCARE in the preceding tables, this pattern is consistent with the HAR-RQ-RES benchmark model assigning too little probability to large negative tail events and hence potentially underestimating tail risk.
    RTCARE forecasts show sharper, short-lived responses to large returns and volatility episodes, whereas HAR-RQ-RES forecasts adjust more gradually, reflecting differences in the dynamic specifications and how realized tail measures enter the models.
    \newpage
    \section{Conclusion}
    This paper proposes RTCARE models, a family of joint VaR--ES models that incorporate RQ and RES into autoregressive expectile dynamics.
    In the numerical examples, three RTCARE specifications are compared with ten competing models at four quantile levels using the FZ0 loss, the $D$ diagnostic, the violation rate, and forecast-error measures. The RTCARE specifications remain competitive across the tail levels.
    In the application to U.S. stock returns, the proposed models and competing benchmarks are evaluated in-sample and in one-day-ahead out-of-sample forecasting, together with ES backtests and model confidence set analysis. The proposed class delivers accurate coverage, strong ES calibration, and competitive out-of-sample forecast performance without relying on a fully parametric return distribution. 
    Further work can examine the noise and bias of realized quantiles and realized ES measures, and can extend the
    expectile dynamics to capture additional effects, such as leverage effects associated with realized volatility measures.

    \section*{Acknowledgments}
    This work is partially supported by JSPS KAKENHI (Grant number 24H00142).
    The computational results are obtained using Rcpp. Generative AI tools were used to assist with coding and proofreading.

\clearpage
\begingroup

\newcommand{\diagnostictablewidth}{0.80\textwidth}
\newcommand{\rankingtablewidth}{0.90\textwidth}
\newcommand{\backtesttablewidth}{0.90\textwidth}
\newcommand{\oosbacktesttablewidth}{0.82\textwidth}
\newcommand{\mcstablewidth}{0.80\textwidth}
\renewcommand{\passmark}{\ensuremath{^{*}}}
\renewcommand{\mcsmark}{\ensuremath{\circ}}
\captionsetup[table]{font=tablecaption,skip=3pt}
\captionsetup[figure]{font=tablecaption}
\AtBeginEnvironment{table}{%
  \small
  \renewcommand{\arraystretch}{1.05}%
}
\spacingset{1}
\appendix

\begin{center}
{\Large
{\bf Supplementary Material for \\``Realized Tail-Risk Conditional Autoregressive Expectile Models: Accurate and Stable Forecasting Performance''}
}
\end{center}

\vspace{0.5cm}
\begin{center}
{\large Keisuke Sasaki$^1$ and Yasuhiro Omori$^2$}

\medskip
$^1$Graduate School of Economics, University of Tokyo, Tokyo 113-0033, Japan, \\
$^2$Faculty of Economics, University of Tokyo, Tokyo 113-0033, Japan \\
\end{center}

\vspace{0.5cm}

\section{Construction of RQ and RES}
\label{supp:rq-res-construction}

High-frequency intraday returns are converted into daily realized tail-risk
measures using the business-time sampling construction of
\citet{Dimitriadis22} and \citet{DongTse17}. The construction used in the
paper is given below for completeness.

\noindent
{\it Step 1. Construct a realized-volatility profile.}
Let $t$ index trading days and let $i$ index intraday observations within day
$t$, with $r_{t,i}$ denoting the corresponding intraday log return. We use the
jump-robust tripower construction of \citet{BarndorffNielsen06}. Its
tripower-variation contribution is
\begin{align}
    RV_{t,i} =
    \left(2^{1/3}\frac{\Gamma(5/6)}{\Gamma(1/2)}\right)^{-3}
    |r_{t,i}|^{2/3}|r_{t,i+1}|^{2/3}|r_{t,i+2}|^{2/3},
\end{align}
where the three returns are consecutive observations on the preliminary
intraday grid. Cumulative sums of these contributions define the
intraday volatility profile.

\noindent
{\it Step 2. Construct business time.}
Let $m_t$ be the number of tripower contributions on day $t$. The normalized
cumulative-volatility map is
\begin{align}
    \theta_t(k) =
    \frac{\sum_{i=1}^{k} RV_{t,i}}
         {\sum_{i=1}^{m_t} RV_{t,i}},
    \qquad k=1,\ldots,m_t,
\end{align}
with $\theta_t(0)=0$. This maps calendar time to cumulative business time on
$[0,1]$.

\noindent
{\it Step 3. Map business time back to calendar time.}
Let $c_t$ denote the number of target intervals retained on day $t$. The
implementation determines this count from the official session length and the
selected target interval. Shortened sessions can therefore have a different
$c_t$, and sessions for which the requested number of returns cannot be
obtained are excluded. For $u_j=j/c_t$, $j=0,\ldots,c_t$, a monotone
cubic-spline interpolation of $\theta_t$ is used to obtain the inverse times
$\theta_t^{-1}(u_j)$. The last available price at each time is retained and
denoted by $p^b_{t,j}$.

\noindent
{\it Step 4. Form business-time returns.}
The prices sampled at equally spaced points in business time produce
\begin{align*}
    r^b_{t,j}=\log p^b_{t,j}-\log p^b_{t,j-1},
    \qquad j=1,\ldots,c_t.
\end{align*}

\noindent
{\it Step 5. Compute RQ and RES.}
Let $\widehat Q_{t,\alpha}$ be the empirical lower-tail $\alpha$-quantile of
the $c_t$ business-time returns, evaluated with the type-7 sample-quantile
convention and linear interpolation. The realized quantile is scaled to the
daily frequency as
\begin{align}
    RQ_t(\alpha)=c_t^{1/2}\widehat Q_{t,\alpha}.
\end{align}
Writing $\widehat{\mathbb{E}}_t[\cdot\mid\cdot]$ for the empirical conditional
mean over the business-time returns, the realized ES is
\begin{align}
    RES_t(\alpha)=c_t^{1/2}
    \widehat{\mathbb{E}}_t\!\left[
    r^b_{t,j}\,\middle|\,r^b_{t,j}\leq\widehat Q_{t,\alpha}\right].
\end{align}

\section{MCMC Sampling Scheme for RTCARE}
\label{appendix:mcmc-algorithm}
This section describes the MCMC algorithm for the RTCARE model. The expectile level $\tau$ is fixed within one MCMC run. Write
$x_t=RQ_t$, $y=(y_1,\ldots,y_n)'$, and $x=(x_1,\ldots,x_n)'$, where $n$ is the estimation sample size. Treat the initial expectile $\mu_0$ as fixed.
\subsection{Pseudo-Likelihood and Priors}
The pseudo-likelihood is
\begin{align}
 L(\beta_0,\beta_1,\beta_2,\alpha_q,\beta_q,\sigma_r^2,\sigma_u^2
 \mid y,x,\tau)
 &\propto (\sigma_r^2)^{-n/2}
 \exp\left\{-\frac{1}{\sigma_r^2}
 \sum_{t=1}^n\left|\tau-I(y_t<\mu_t)\right|(y_t-\mu_t)^2\right\}
    \notag\\
 &\quad\times (\sigma_u^2)^{-n/2}
 \exp\left\{-\frac{1}{2\sigma_u^2}
 \sum_{t=1}^n(x_t-\alpha_q-\beta_q\mu_t)^2\right\},
 \label{eq:mcmc-likelihood}
\end{align}
where $\mu_t$ is recomputed from the RTCARE recursion defined in the main manuscript whenever
$(\beta_0,\beta_1,\beta_2)$ changes.

Before imposing the coefficient restriction, the independent priors are
\begin{align}
    \beta_j &\sim \N(m_{\beta j},s_{\beta j}^2), &&j=0,1,2,\\
    \alpha_q &\sim \N(m_{\alpha q}^{0},s_{\alpha q}^2),
    & \beta_q &\sim \N(m_{\beta q},s_{\beta q}^2),\\
    \sigma_r^2 &\sim \IG(a_r,b_r),
    & \sigma_u^2 &\sim \IG(a_u,b_u),
\end{align}
where the inverse-gamma distribution is parameterized by
\[
    p(x)\propto x^{-(a+1)}\exp(-b/x).
\]
For the baseline specification, the hyperparameters are set as follows:
\begin{align}
 (m_{\beta0},m_{\beta1},m_{\beta2},m_{\alpha q}^{0},m_{\beta q})
 &=(0,0,0,0,0),\\
 (s_{\beta0}^2,s_{\beta1}^2,s_{\beta2}^2,s_{\alpha q}^2,s_{\beta q}^2)
 &=(10,10,10,10,10),\\
(a_r,b_r,a_u,b_u)&=(0.02,0.02,0.02,0.02).
\end{align}
\subsection{Joint Posterior Distribution}
Combining the pseudo-likelihood and the prior distributions gives the joint posterior
density on the support $\sigma_r^2>0$, $\sigma_u^2>0$, and $|\beta_1+\beta_2\beta_q|<1$:
\begin{align}
&\pi(\beta_0,\beta_1,\beta_2,\alpha_q,\beta_q,\sigma_r^2,\sigma_u^2
\mid y,x,\tau)
\notag\\
&\quad\propto
(\sigma_r^2)^{-(a_r+1+n/2)}
\exp\left\{-\frac{b_r+\sum_{t=1}^n
\left|\tau-I(y_t<\mu_t)\right|(y_t-\mu_t)^2}{\sigma_r^2}\right\}
\notag\\
&\qquad\times
(\sigma_u^2)^{-(a_u+1+n/2)}
\exp\left\{-\frac{b_u+\frac{1}{2}\sum_{t=1}^n
(x_t-\alpha_q-\beta_q\mu_t)^2}{\sigma_u^2}\right\}
\notag\\
&\qquad\times
\exp\left\{-\frac{1}{2}\sum_{j=0}^{2}
\frac{(\beta_j-m_{\beta j})^2}{s_{\beta j}^2}
-\frac{(\alpha_q-m_{\alpha q}^{0})^2}{2s_{\alpha q}^2}
-\frac{(\beta_q-m_{\beta q})^2}{2s_{\beta q}^2}\right\}
\label{eq:joint-posterior}
\end{align}
The posterior density is zero outside this support.

\subsection{Full Conditional Posterior Distributions}

\subsubsection*{Generation of $(\beta_0,\beta_1,\beta_2,\beta_q)$}

The intercept $\alpha_q$ is integrated out for the joint update of
$(\beta_0,\beta_1,\beta_2,\beta_q)$. This block is updated by a joint
random-walk Metropolis--Hastings step. Propose
\begin{align}
 (\beta_0^\star,\beta_1^\star,\beta_2^\star,\beta_q^\star)'
 \sim\N_4\left((\beta_0,\beta_1,\beta_2,\beta_q)',s_j^2\Sigma_j\right),
\end{align}
Reject any proposal that violates $|\beta_1^\star+\beta_2^\star\beta_q^\star|<1$. Otherwise, accept it with the Metropolis--Hastings probability computed from the posterior with $\alpha_q$ integrated out. Recompute
$\mu_{1:n}$ using the accepted values of $(\beta_0,\beta_1,\beta_2)$.

\subsubsection*{Generation of $\alpha_q$}

The full conditional posterior distribution is
\begin{align}
 \alpha_q\mid\beta_0,\beta_1,\beta_2,\beta_q,\sigma_u^2,y,x,\tau
 &\sim \N\left(
 \frac{m_{\alpha q}^{0}/s_{\alpha q}^2
 +\sigma_u^{-2}\sum_{t=1}^n(x_t-\beta_q\mu_t)}
 {s_{\alpha q}^{-2}+n\sigma_u^{-2}},
 \left(s_{\alpha q}^{-2}+n\sigma_u^{-2}\right)^{-1}
 \right).
\end{align}

\subsubsection*{Generation of $\sigma_r^2$}

The full conditional posterior distribution is
\begin{align}
 \sigma_r^2\mid\beta_0,\beta_1,\beta_2,y,x,\tau
 &\sim \IG\left(a_r+\frac{n}{2},
 b_r+\sum_{t=1}^n\left|\tau-I(y_t<\mu_t)\right|(y_t-\mu_t)^2\right).
\end{align}

\subsubsection*{Generation of $\sigma_u^2$}

The full conditional posterior distribution is
\begin{align}
 \sigma_u^2\mid\alpha_q,\beta_q,\mu_{1:n},x
 &\sim \IG\left(a_u+\frac{n}{2},
 b_u+\frac{1}{2}\sum_{t=1}^n
 (x_t-\alpha_q-\beta_q\mu_t)^2\right).
\end{align}

\subsection{MCMC Algorithm}

The MCMC algorithm proceeds as follows.
\begin{enumerate}
 \item Initialize
 $(\beta_0,\beta_1,\beta_2,\alpha_q,\beta_q,\sigma_r^2,\sigma_u^2)$ within the parameter support.
 \item Sample $(\alpha_q,\beta_0,\beta_1,\beta_2,\beta_q)\mid
 \sigma_r^2,\sigma_u^2,y,x,\tau$ as follows.
 \begin{enumerate}
  \item Sample $(\beta_0,\beta_1,\beta_2,\beta_q)\mid
  \sigma_r^2,\sigma_u^2,y,x,\tau$ by the joint MH update above.
  \item Sample $\alpha_q\mid\beta_0,\beta_1,\beta_2,\beta_q,
  \sigma_u^2,y,x,\tau$ from its normal full conditional distribution.
 \end{enumerate}
 \item Sample $\sigma_r^2\mid\beta_0,\beta_1,\beta_2,y,x,\tau$.
 \item Sample $\sigma_u^2\mid\alpha_q,\beta_q,\mu_{1:n},x$.
 \item Repeat Steps 2--4 for a sufficiently large number of iterations.
\end{enumerate}

\subsection{Forecasting from the MCMC Draws}

After burn-in, retained draws are used to construct the one-step-ahead
forecasts as follows.

For these forecasting formulas, we impose the zero conditional mean assumption
$\mathbb{E}(y_t\mid\mathcal{F}_{t-1})=0$. Hence, the expected-return term in the
general VaR--ES mapping vanishes.

\begin{align}
    \mu_{n+1}&=\beta_0+\beta_1\mu_n+\beta_2RQ_n,\\
    VaR_{n+1,\alpha}&=\mu_{n+1},\\
    ES_{n+1,\alpha}&=\left(1+\frac{\tau}{(1-2\tau)\alpha}\right)\mu_{n+1}.
\end{align}
Posterior means, medians, and credible intervals are calculated from the
retained forecast draws.

\section{Ten Competing Forecast Models}
\label{appendix:competing-models}

This section of the Supplementary Material gives the specifications of the ten competing models used in the numerical examples and in the rolling out-of-sample study. For the in-sample comparison, the HAR-RQ-RES benchmark is replaced by the contemporaneous RQ-RES benchmark. For any conditional scale model with $y_t=\sigma_t\varepsilon_t$, write $q_{\varepsilon,\alpha}$ and
$e_{\varepsilon,\alpha}$ for the $\alpha$-quantile and lower-tail mean of
$\varepsilon_t$. The corresponding forecasts are
$VaR_{t,\alpha}=\sigma_tq_{\varepsilon,\alpha}$ and
$ES_{t,\alpha}=\sigma_te_{\varepsilon,\alpha}$.
The equations below define the specifications used in the comparison.

\subsection{CARE RV Family Models}

All CARE-type competitors use
\begin{equation}
 y_t=\mu_t+\varepsilon_t,\qquad
 \varepsilon_t\sim\AN(0,\sigma_r^2,\tau),
\end{equation}
and map the fitted expectile to VaR and ES through
the VaR--ES mapping defined in the main manuscript.
Here $RV_t$ denotes realized variance.

\paragraph{CARE RV model}
\begin{align}
 \mu_t&=\beta_0+\beta_1\mu_{t-1}+\beta_2RV_{t-1},\\
 RV_t&=\alpha_v+\beta_v\mu_t+u_t, \quad \quad u_t\sim\N(0,\sigma_u^2).
\end{align}

\paragraph{CARE RV-RQ model}
\begin{equation}
 \mu_t=\beta_0+\beta_1\mu_{t-1}+\beta_2RV_{t-1}+\beta_3RQ_{t-1}.
\end{equation}

\paragraph{CARE RV-RQ-RES model}
\begin{equation}
 \mu_t=\beta_0+\beta_1\mu_{t-1}+\beta_2RV_{t-1}
 +\beta_3RQ_{t-1}+\beta_4RES_{t-1}.
\end{equation}

\paragraph{Realized CARE-SAV model}
\begin{align}
 \mu_t&=\beta_0+\beta_1\mu_{t-1}+\beta_2\sqrt{RV_{t-1}},\\
 \sqrt{RV_t}&=\xi+\phi|\mu_t|+\delta_1z_t+\delta_2(z_t^2-\overline{z^2})+u_t,
 \quad \quad u_t\sim\N(0,\sigma_u^2),\\
 z_t&=y_t/\mu_t, \quad \overline{z^2}=n^{-1}\sum_{t=1}^nz_t^2.
\end{align}

\subsection{One-Factor GAS Model}

The one-factor GAS model uses the dynamic factor $f_t$.
\begin{align}
 y_t&=\exp(f_t)\eta_t,\\
 q_t&=a\exp(f_t), \quad e_t=b\exp(f_t),\\
 f_t&=\omega+\beta f_{t-1}
 +\frac{\gamma}{e_{t-1}}
 \left\{\frac{I(y_{t-1}\leq q_{t-1})y_{t-1}}{\alpha}-e_{t-1}\right\}.
\end{align}

\subsection{GARCH Family Models}

\paragraph{GARCH(1,1) model}
\begin{align}
 y_t&=\sigma_t\varepsilon_t, \quad \quad \varepsilon_t \sim\N(0,1),\\
 \sigma_t^2&=\omega+a y_{t-1}^2+b\sigma_{t-1}^2.
\end{align}
Here $q_{\varepsilon,\alpha}=\Phi^{-1}(\alpha)$ and
$e_{\varepsilon,\alpha}=-\phi\{\Phi^{-1}(\alpha)\}/\alpha$.

\paragraph{GARCH-t(1,1) model}
The conditional-scale recursion is the same as above, but
\begin{equation}
 \varepsilon_t\sim t_\nu(0,1),
\end{equation}
where $t_\nu(0,1)$ is standardized to have unit variance. Its quantile and
lower-tail mean replace the Gaussian $q_{\varepsilon,\alpha}$ and
$e_{\varepsilon,\alpha}$.

\paragraph{Realized GARCH(1,1) model}
\begin{align}
 y_t&=\sqrt{h_t}\varepsilon_t, &\varepsilon_t&\sim\N(0,1),\\
 \log h_t&=\omega+\beta\log h_{t-1}+\gamma\log RV_{t-1},\\
 \log RV_t&=\xi+\phi\log h_t+\tau(\varepsilon_t)+u_t, &u_t&\sim\N(0,\sigma_u^2),\\
 \tau(\varepsilon_t)&=\eta_1\varepsilon_t+\eta_2(\varepsilon_t^2-1).
\end{align}
VaR and ES use the Gaussian mapping with conditional scale $\sqrt{h_t}$.

\paragraph{Realized GARCH-t(1,1) model}
This model retains the Realized GARCH state and measurement equations above,
but sets $\varepsilon_t\sim t_\nu(0,1)$ and uses the standardized Student-$t$ quantile
and lower-tail mean in the VaR--ES mapping.

\subsection{HAR-RQ-RES Model}

Define weekly and monthly averages by
\begin{align}
 \overline{RQ}^{(w)}_t&=\frac{1}{5}\sum_{j=0}^{4}RQ_{t-j}, &
 \overline{RQ}^{(m)}_t&=\frac{1}{22}\sum_{j=0}^{21}RQ_{t-j},\\
 \overline{RES}^{(w)}_t&=\frac{1}{5}\sum_{j=0}^{4}RES_{t-j}, &
 \overline{RES}^{(m)}_t&=\frac{1}{22}\sum_{j=0}^{21}RES_{t-j}.
\end{align}
The benchmark consists of the following two equations, which are estimated separately:
\begin{align}
 RQ_{t+1}&=c_q+b_{q,d}RQ_t+b_{q,w}\overline{RQ}^{(w)}_t
 +b_{q,m}\overline{RQ}^{(m)}_t+u_{q,t+1},\\
 RES_{t+1}&=c_e+b_{e,d}RES_t+b_{e,w}\overline{RES}^{(w)}_t
 +b_{e,m}\overline{RES}^{(m)}_t+u_{e,t+1}.
\end{align}
It uses $\widehat{VaR}_{t+1|t}=\widehat{RQ}_{t+1|t}$ and
$\widehat{ES}_{t+1|t}=\widehat{RES}_{t+1|t}$.
The two equations are estimated separately by MCMC.

\section{Tables and Plots of RMSE in Numerical Examples}
\label{appendix:numerical-figures}

\begingroup
\spacingset{1}
   The main text focuses on the FZ0 loss, the $D$ diagnostics, and violation rates.
   Tables~\ref{tab:sim-gjr-rmse-var} and~\ref{tab:sim-gjr-rmse-es} report the
   replication means and standard deviations of the VaR and ES RMSE measures.
   They complement the corresponding boxplots below.
   Because the data-generating process supplies the true conditional VaR and ES paths, the numerical design also permits RMSE comparisons. Figures
    \ref{fig:numerical-rmse-var} and \ref{fig:numerical-rmse-es} report the
   corresponding replication distributions.
   The RMSE measures the distance between a model's forecast tail quantity and the corresponding true conditional quantity implied by the data-generating process, so smaller values indicate more accurate recovery.  Each boxplot summarizes the 50 replications.

\begin{table}[htbp]
\centering
\caption{Numerical example. VaR RMSE across the four tail levels. Entries are replication means with standard deviations in parentheses. * denotes the smallest entry in each column.}
\label{tab:sim-gjr-rmse-var}

\fontsize{10}{11}\selectfont
\setlength{\tabcolsep}{3pt}
\renewcommand{\arraystretch}{0.88}
\begin{tabular*}{\textwidth}{@{\extracolsep{\fill}}lrrrr}
\toprule
Model & $5\%$ & $2.5\%$ & $1\%$ & $0.5\%$ \\
\midrule
RTCARE & 0.548 (0.328) & 0.766 (0.540) & 1.155 (0.828) & 1.535 (0.938) \\
RTCARE NO INT & 0.470 (0.281) & *0.634 (0.415) & 0.972 (0.604) & 1.307 (0.700) \\
RTCARE-RES & 0.563 (0.339) & 0.782 (0.557) & 1.139 (0.776) & 1.523 (0.929) \\
\midrule
CARE RV & 0.950 (1.099) & 1.241 (1.491) & 1.653 (1.780) & 2.035 (1.912) \\
CARE RV-RQ & 0.898 (0.927) & 1.202 (1.084) & 1.766 (1.370) & 2.346 (1.687) \\
CARE RV-RQ-RES & 0.957 (0.909) & 1.265 (1.002) & 1.849 (1.379) & 2.259 (1.594) \\
Realized CARE-SAV & 0.543 (0.297) & 0.741 (0.479) & 1.092 (0.650) & 1.444 (0.775) \\
\midrule
One-factor GAS & 0.596 (0.307) & 1.039 (0.662) & 2.311 (1.878) & 3.399 (3.021) \\
\midrule
GARCH & *0.400 (0.125) & 0.708 (0.203) & 1.298 (0.360) & 1.877 (0.518) \\
GARCH-t & 0.472 (0.149) & 0.655 (0.192) & *0.908 (0.248) & *1.113 (0.299) \\
Realized GARCH & 0.471 (0.199) & 0.739 (0.204) & 1.283 (0.289) & 1.837 (0.408) \\
Realized GARCH-t & 0.502 (0.154) & 0.677 (0.191) & 0.918 (0.244) & 1.115 (0.296) \\
\midrule
HAR-RQ-RES & 0.435 (0.149) & 0.809 (0.253) & 1.473 (0.448) & 1.910 (0.565) \\
\bottomrule
\end{tabular*}
\end{table}

\begin{table}[htbp]
\centering
\caption{Numerical example. ES RMSE across the four tail levels. Entries are replication means with standard deviations in parentheses. * denotes the smallest entry in each column.}
\label{tab:sim-gjr-rmse-es}

\fontsize{10}{11}\selectfont
\setlength{\tabcolsep}{3pt}
\renewcommand{\arraystretch}{0.88}
\begin{tabular*}{\textwidth}{@{\extracolsep{\fill}}lrrrr}
\toprule
Model & $5\%$ & $2.5\%$ & $1\%$ & $0.5\%$ \\
\midrule
RTCARE & 0.823 (0.531) & 1.076 (0.672) & 1.550 (0.991) & 2.070 (1.192) \\
RTCARE NO INT & *0.675 (0.385) & *0.872 (0.487) & 1.259 (0.739) & 1.713 (1.010) \\
RTCARE-RES & 0.854 (0.561) & 1.098 (0.688) & 1.545 (0.983) & 2.042 (1.207) \\
\midrule
CARE RV & 1.375 (1.639) & 1.672 (1.891) & 2.137 (2.203) & 2.563 (2.280) \\
CARE RV-RQ & 1.277 (1.306) & 1.584 (1.348) & 2.198 (1.531) & 2.898 (1.869) \\
CARE RV-RQ-RES & 1.357 (1.328) & 1.669 (1.211) & 2.280 (1.576) & 2.853 (1.908) \\
Realized CARE-SAV & 0.838 (0.516) & 1.066 (0.634) & 1.495 (0.878) & 1.936 (1.002) \\
\midrule
One-factor GAS & 0.860 (0.446) & 1.416 (0.746) & 2.854 (2.132) & 3.965 (3.139) \\
\midrule
GARCH & 0.985 (0.275) & 1.480 (0.409) & 2.311 (0.638) & 3.086 (0.855) \\
GARCH-t & 0.746 (0.207) & 0.945 (0.255) & *1.227 (0.332) & *1.463 (0.407) \\
Realized GARCH & 0.986 (0.231) & 1.453 (0.323) & 2.257 (0.507) & 3.015 (0.693) \\
Realized GARCH-t & 0.763 (0.209) & 0.954 (0.256) & 1.231 (0.338) & 1.473 (0.423) \\
\midrule
HAR-RQ-RES & 1.109 (0.332) & 1.734 (0.510) & 2.774 (0.807) & 3.662 (1.055) \\
\bottomrule
\end{tabular*}
\end{table}

   For VaR, Figure~\ref{fig:numerical-rmse-var} shows that the GARCH specifications are often among the most accurate at the less extreme tail levels, while the RTCARE specifications remain competitive and display comparatively compact boxes.  The CARE RV model variants have wider boxes and longer upper whiskers, especially at 1\% and 0.5\%, indicating greater replication-to-replication instability.  The RMSE generally rises as the tail level becomes more extreme, reflecting the smaller effective number of tail observations.  The GARCH models are competitive in terms of both average RMSE and variability across replications. The three RTCARE models exhibit less variability than the CARE RV variants.

   For ES, Figure~\ref{fig:numerical-rmse-es} gives a similar but more pronounced pattern.  The GARCH-t and the RTCARE specifications based on the realized tail measures tend to have the smallest or near-smallest central errors at the moderate tail levels.  In contrast, the ES boxes for the HAR-RQ-RES model and the CARE RV model variants become noticeably larger or wider and extend further upward at 1\% and 0.5\%, showing greater variability in their ES forecast errors across replications. 
   This widening matters because ES depends on losses beyond the VaR threshold, not only on its location. These RMSE results complement the results of the FZ0 losses, $D$ diagnostics, and violation rates in the main text.
   The three RTCARE models do not attain the lowest RMSE at every tail level, but they generally have smaller VaR and ES RMSE values than the CARE RV variants as the tail becomes more extreme.
\endgroup
\begin{figure}[htbp]
  \centering
  \makebox[\textwidth][c]{%
    \includegraphics[width=\textwidth,height=0.78\textheight,keepaspectratio]{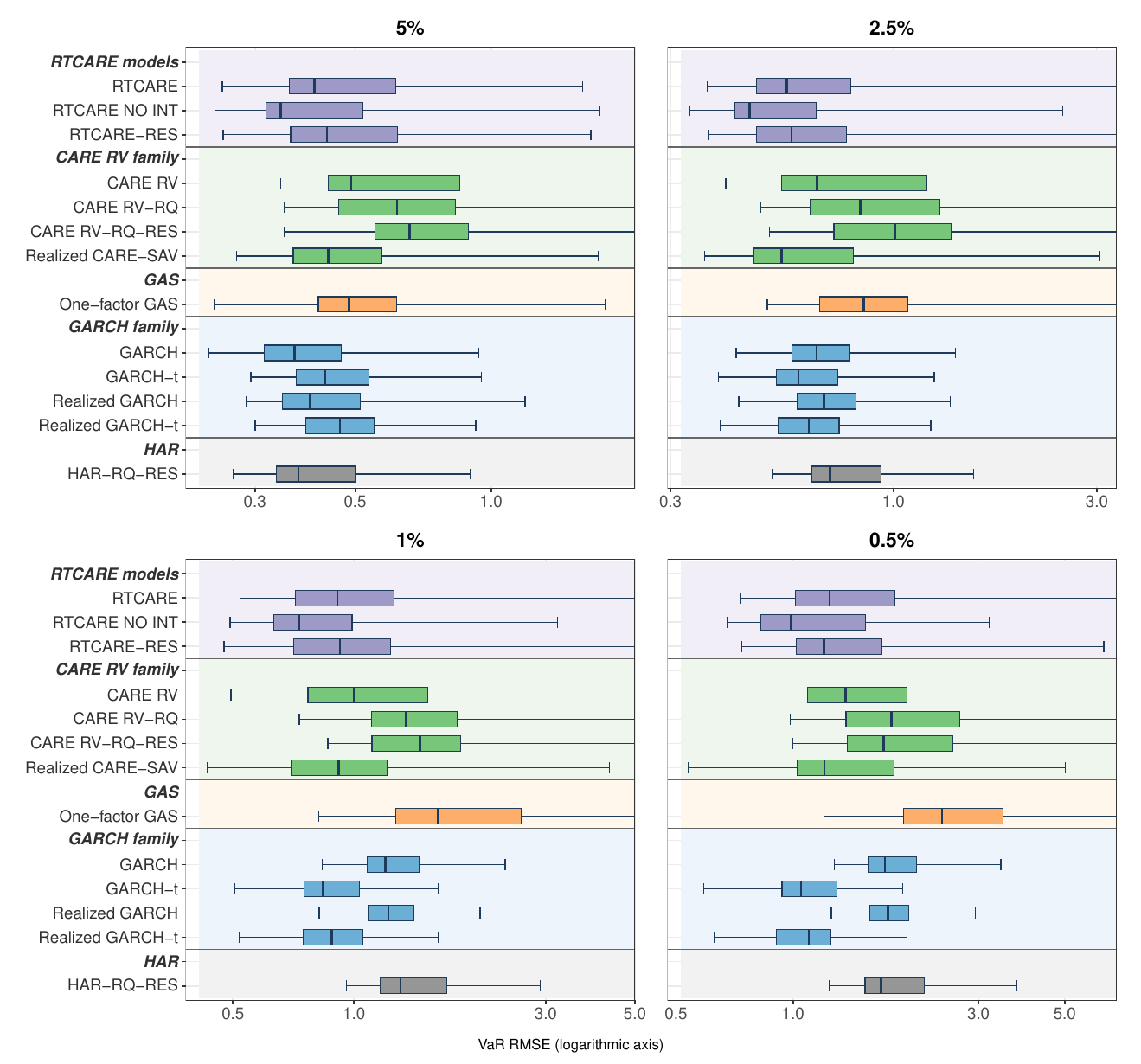}}
  \caption{Boxplots of RMSE for VaR across the 50 replications in the numerical example.}
  \label{fig:numerical-rmse-var}
\end{figure}
\begin{figure}[htbp]
  \centering
  \makebox[\textwidth][c]{%
    \includegraphics[width=\textwidth,height=0.78\textheight,keepaspectratio]{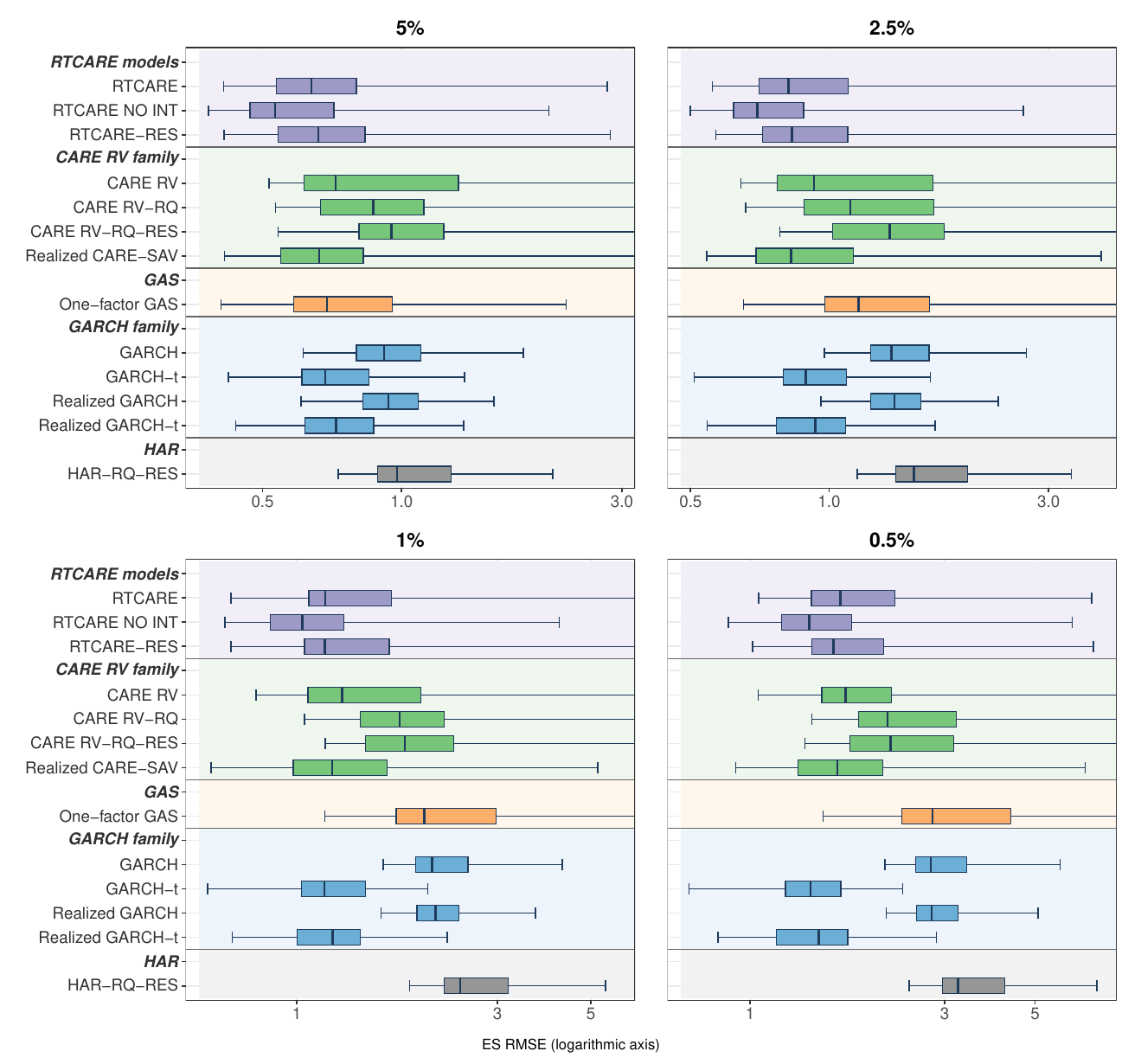}}
  \caption{Boxplots of RMSE for ES across the 50 replications in the numerical example.}
  \label{fig:numerical-rmse-es}
  \vspace{5mm}
\end{figure}
\newpage
\section{Time-Series Plots of Daily Returns, RQ, and RES}
\label{supp:additional-empirical-figure}
The time-series plot in Figure~\ref{fig:supp-empirical-daily-rq-res} illustrates the daily return series together with the realized
quantile (RQ) and realized expected shortfall (RES) measures used in the
empirical analysis. These measures are constructed from three-minute
intraday returns at the 5\% tail level and are plotted over the full
2005--2024 sample. RQ tracks the lower-tail quantile of daily returns,
whereas RES summarizes the average return below that threshold and is
therefore typically no greater than RQ. The plot shows how the two realized
tail-risk measures vary over time and provides a visual description of the
inputs used in the empirical models.
\begin{figure}[htbp]
  \centering
  \includegraphics[width=\textwidth]{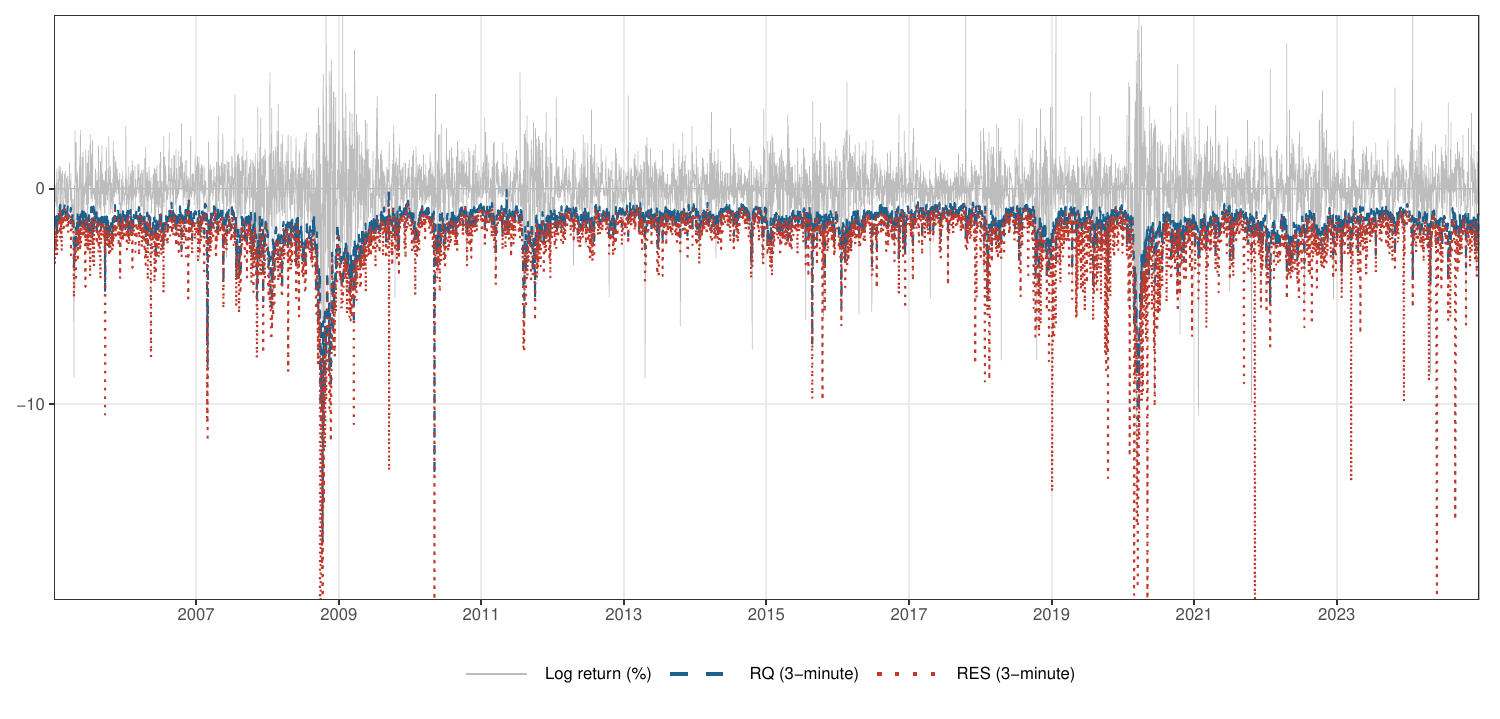}
  \caption{IBM daily returns and daily RQ and RES at the 5\% tail level, constructed from three-minute intraday returns, over the full 2005--2024 sample.}
  \label{fig:supp-empirical-daily-rq-res}
\end{figure}

\section{Ranking Plots}
\label{appendix:ranking}
This section reports the rank plots omitted from the main text. The ranks are
computed separately for the FZ0 loss, the $D$
diagnostic, and the violation rate at each tail level, with Rank 1 denoting
the best-performing model for the corresponding criterion. The ranks use
unrounded values, with exact ties assigned the minimum rank. For violation
rates, models are ranked by the absolute deviation from the nominal level.
\subsection{In-Sample Ranking Summary}
Figure~\ref{fig:is_ranking_dotplot} shows the in-sample ranks based on the FZ0 loss, $D$ diagnostics, and violation rates reported in the main
text. The RQ-RES benchmark takes the top rank for FZ0 losses, but its
advantage does not carry over uniformly to the $D$ and violation-rate
criteria. Among the dynamic models, the RTCARE variants remain in the
leading tier, while the GARCH and Realized GARCH models move toward the
bottom as the tail becomes more extreme.
\begin{figure}[htbp]
  \centering
  \makebox[\textwidth][c]{%
    \includegraphics[width=\textwidth]{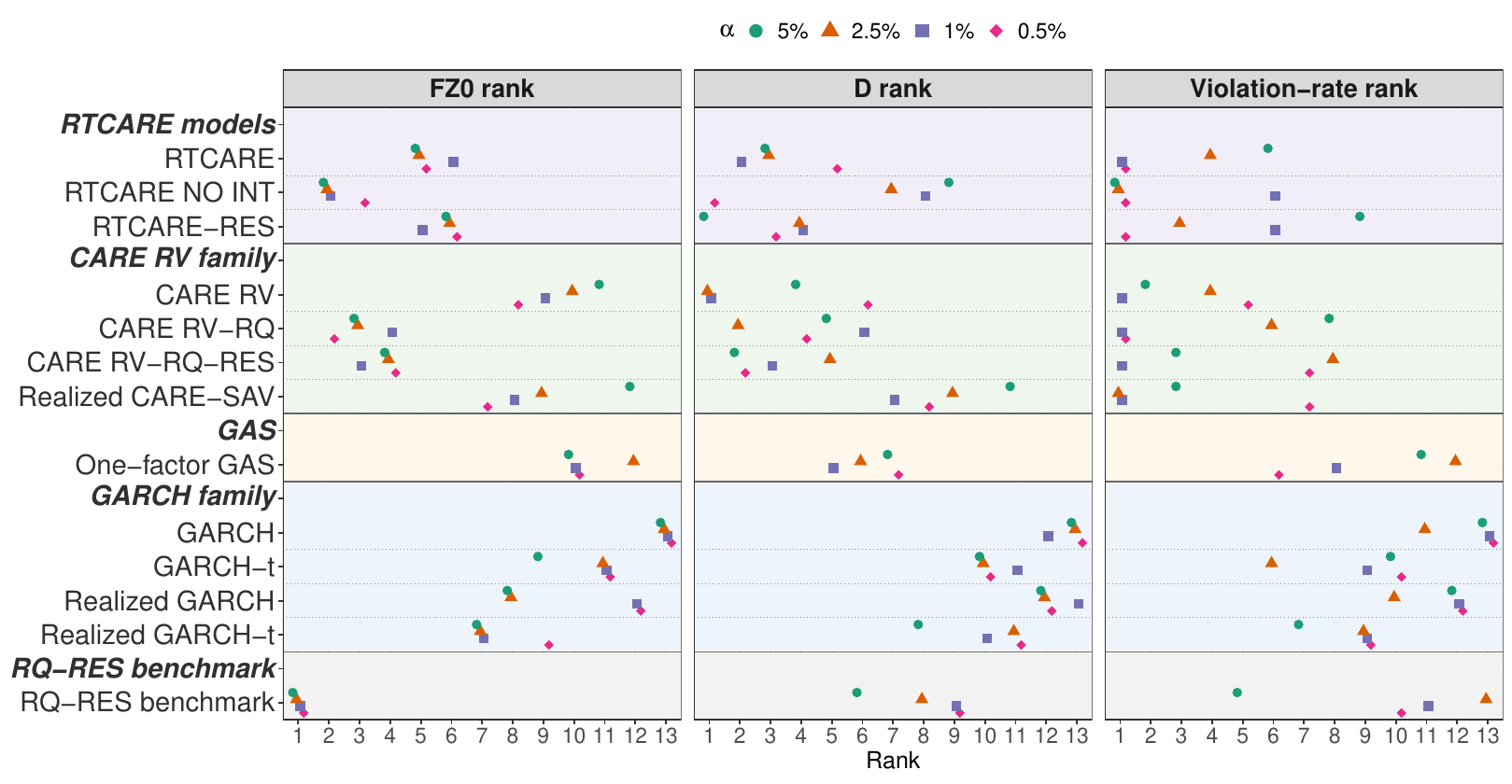}}
  \caption{Rank plot for the FZ0 loss, $D$ diagnostics, and violation rates
  (in-sample). Each dot represents a specific quantile level $\alpha$.}
  \label{fig:is_ranking_dotplot}
\end{figure}
\subsection{Out-of-Sample Ranking Summary}
Figure~\ref{fig:oos_ranking_dotplot} shows the out-of-sample ranks based on
the FZ0 loss, $D$ diagnostics, and violation rates reported in the main
text. The three RTCARE specifications occupy the leading FZ0 positions at all
four levels and remain competitive with respect to the $D$ diagnostics and
violation rates. Their relative advantage is clearest in the extreme tails,
where the GARCH and Realized GARCH models show larger losses and $D$
values. Realized CARE-SAV is competitive on $D$ diagnostics and
violation rates, whereas HAR-RQ-RES is consistently weaker across the three
measures.
\begin{figure}[htbp]
  \centering
  \makebox[\textwidth][c]{%
    \includegraphics[width=\textwidth]{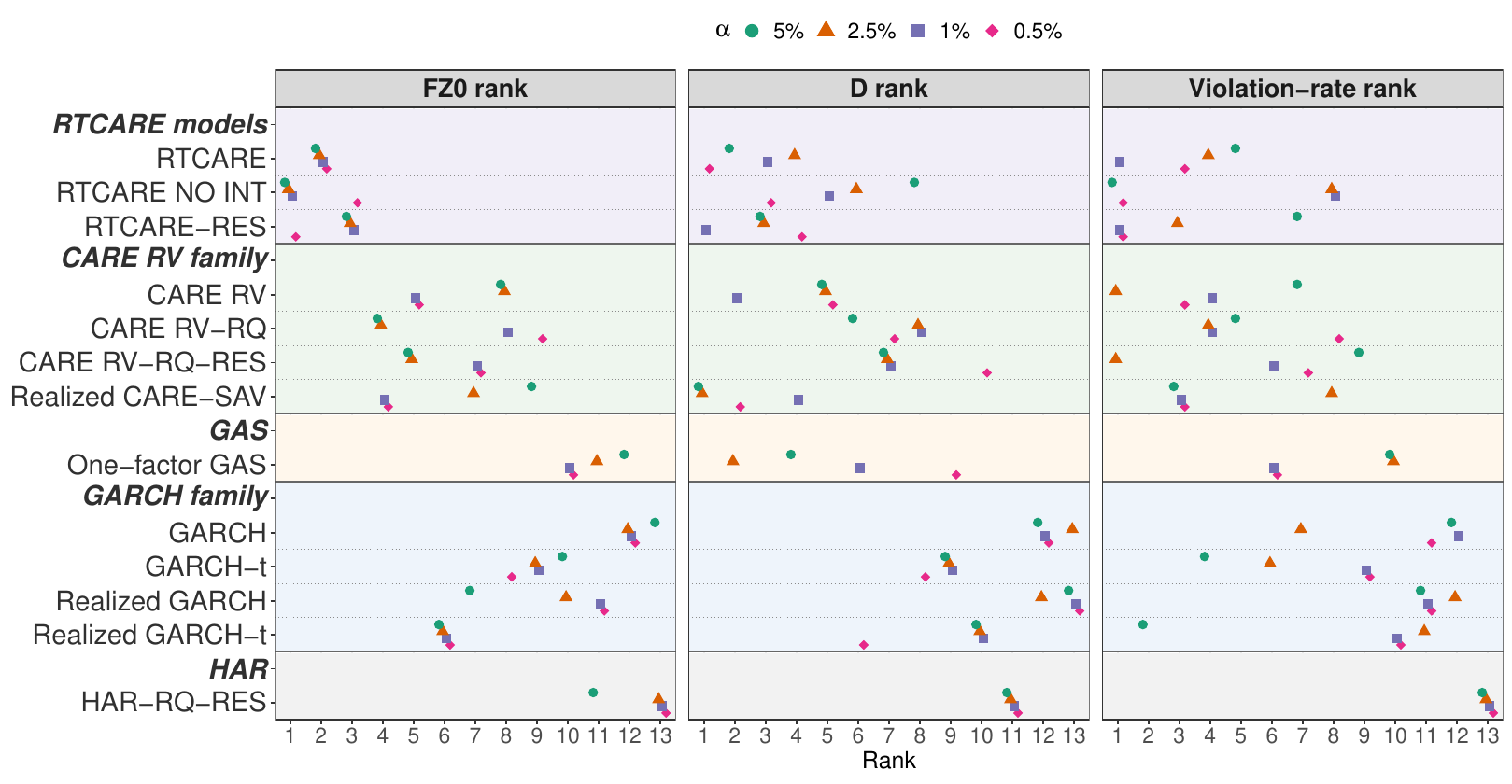}}
  \caption{Rank plot for the FZ0 loss, $D$ diagnostics, and violation rates
  (out-of-sample). Each dot represents a specific quantile level $\alpha$.}
  \label{fig:oos_ranking_dotplot}
  \vspace{5mm}
\end{figure}

\section{Regression-Based ES Backtesting} 
    \label{appendix:backtest}
    Let $\widehat q_t$ and $\widehat e_t$ denote the reported VaR and ES
    estimates. The tests are based on a joint VaR--ES regression framework.
    Under correct calibration, the conditional quantile of the
    quantile residual is zero and the conditional ES of the ES residual is zero.
    The joint regression parameters are estimated by M-estimation using a
    strictly consistent loss function for the pair of VaR and ES. See
    \citet{BayerDimitriadis22} for details. 

    \begin{enumerate}
    \item \emph{Strict test}. The test uses only $\hat{e}_t$ as an explanatory variable in both equations of the joint regression:
    \begin{align}
      y_t&=a_q^{(s)}+b_q^{(s)}\widehat e_t+u_t^q,\\
      y_t&=a_s+b_s\widehat e_t+u_t^e,
      \label{eq:esr-strict}
    \end{align}
    where $u_t^q$ and $u_t^e$ are the quantile and ES residuals, respectively. We test the null hypothesis $H_0:a_s=0, b_s=1$, which implies that the ES forecasts are correctly specified.
    
    \item \emph{Auxiliary test}. We use both $\hat{q}_t$ and $\hat{e}_t$. Consider the joint regression 
    \begin{equation}
      y_t=a_q+b_q\widehat q_t+u_t^q,
      \qquad
      y_t=a_e+b_e\widehat e_t+u_t^e,
      \label{eq:esr-auxiliary}
    \end{equation}
    where the first equation is the quantile regression and the second is the ES regression. We test $H_0:a_e=0, b_e=1$, which assesses ES calibration within the joint VaR--ES regression framework. 
    \item \emph{Intercept test}. 
    This test uses $\hat{e}_t$ as the explanatory variable in the quantile
    equation and fixes the ES slope at one ($b_s=1$). The joint regression is
    \begin{align}
      y_t&=a_q^{(i)}+b_q^{(i)}\widehat e_t+u_t^q,\\
      y_t-\widehat e_t&= a_s+u_t^e.
      \label{eq:esr-intercept}
    \end{align}
    The null hypothesis $H_0:a_s=0$ implies that there is no ES-level bias after the unit-slope restriction is imposed.
    \end{enumerate}

\endgroup

\end{document}